\documentclass{jpp}
\usepackage{graphicx}

\usepackage[utf8]{inputenc}
\usepackage{multirow}
\usepackage[T1]{fontenc}
\usepackage{amsmath}
\usepackage{graphicx}
\usepackage{dcolumn}
\usepackage{bm}
\usepackage{booktabs}
\usepackage{siunitx}
\usepackage[utf8]{inputenc}
\usepackage[T1]{fontenc}
\usepackage{mathptmx}
\usepackage{etoolbox}
\usepackage{hhline}
\usepackage{hyperref}
\usepackage{xcolor}
\hypersetup{
    colorlinks = true,
    linkcolor = blue,
    citecolor = blue,
}
\usepackage{booktabs}
\usepackage{siunitx}
\usepackage{tabularx}
\usepackage[justification=justified]{caption}
\usepackage{comment}

\shorttitle{Simulations of Radiatively-Cooled Magnetic Reconnection Driven by Pulsed Power}
\shortauthor{R. Datta, A. Crilly, J. P. Chittenden et al.}

\title{Simulations of Radiatively-Cooled Magnetic Reconnection Driven by Pulsed Power}

\author{
  	Rishabh Datta\aff{1},
    Aidan Crilly\aff{3},
    Jeremy P. Chittenden\aff{3},
    Simran Chowdhry\aff{1}, 
    Katherine Chandler\aff{2},
    Nikita Chaturvedi\aff{3},
    Clayton E. Myers\aff{2},
    William R. Fox\aff{5},
    Stephanie B. Hansen\aff{2},
    Chris A. Jennings\aff{2},
    Hantao Ji\aff{5,6},
    Carolyn C. Kuranz\aff{7},
    Sergey V. Lebedev\aff{3},
    Dmitri A. Uzdensky\aff{4},
    \and Jack D. Hare\aff{1} \corresp{\email{jdhare@mit.edu}}
}

\affiliation{
	\aff{1}Plasma Science and Fusion Center, Massachusetts Institute of Technology, Cambridge MA, USA
	\aff{2}Sandia National Laboratories, Albuquerque NM, USA
	\aff{3}Blackett Laboratory, Imperial College London, London SW7 2BW, UK
    \aff{4}Center for Integrated Plasma Studies, Physics Department, UCB-390, University of Colorado, Boulder, Colorado, USA
    \aff{5} Princeton Plasma Physics Laboratory, Princeton, NJ, USA
    \aff{6} Department of Astrophysical Sciences, Princeton University, Princeton, NJ, USA
    \aff{7} University of Michigan – Ann Arbor, Michigan USA
}

\begin{document}
\newcommand{\ra}[1]{\renewcommand{\arraystretch}{#1}}
\maketitle

\begin{abstract}
Magnetic reconnection is an important process in astrophysical environments, as it re-configures magnetic field topology and converts magnetic energy into thermal and kinetic energy. 
In extreme astrophysical systems, such as black hole coronae and pulsar magnetospheres, radiative cooling modifies the energy partition by radiating away internal energy, which can lead to the radiative collapse of the reconnection layer. In this paper, we perform two- and three-dimensional simulations to model the MARZ (Magnetic Reconnection on Z) experiments, which are designed to access cooling rates in the laboratory necessary to investigate reconnection in a previously unexplored radiatively-cooled regime. These simulations are performed in GORGON, an Eulerian two-temperature resistive magnetohydrodynamic code, which models the experimental geometry comprising two exploding wire arrays driven by  20 MA of current on the Z machine (Sandia National Laboratories). Radiative losses are implemented using non-local thermodynamic equilibrium tables computed using the atomic code Spk, and we probe the effects of radiation transport by implementing both a local radiation loss model and P$_{1/3}$ multi-group radiation transport. The load produces highly collisional, super-Alfvénic ($M_A \approx 1.5$), supersonic $(M_S \approx 4-5)$ strongly driven plasma flows which generate an elongated reconnection layer ($L/\delta \approx 100, \, S_L \approx 400$). The reconnection layer undergoes radiative collapse when the radiative losses exceed the rates of Ohmic and compressional heating ($\tau_{\text{cool}}^{-1}/\tau_{H}^{-1}\approx 100$); this generates a cold strongly compressed current sheet, leading to an accelerated reconnection rate, consistent with theoretical predictions. Finally, the current sheet is also unstable to the plasmoid instability, but the magnetic islands are extinguished by strong radiative cooling before ejection from the layer.

\end{abstract}

\section{Introduction}

\subsection{Radiative reconnection in astrophysical environments}
\label{sec:intro}

Magnetic reconnection is a fundamental process in magnetized plasmas, responsible for the abrupt rearrangement of magnetic field topology, and the violent conversion of magnetic energy into internal and kinetic energy \citep{yamada2010magnetic,zweibel2016perspectives,ji2022magnetic}. Reconnection drives some of the most energetic events in our Universe, including solar flares, coronal mass ejections, and geomagnetic storms in our solar system \citep{parker1963solar,masuda1994loop,yamada2010magnetic}, as well as similar events in the coronae of other stars, in the accretion disks and jets of Young Stellar Objects (YSOs) \citep{goodson1997time,feigelson1999high,benz2010physical}, and in the interstellar medium \citep{zweibel1989magnetic,brandenburg1995effects,lazarian1999reconnection,heitsch2003fast}.


Due to the dissipation of magnetic energy, radiative emission is a key signature of reconnection in many astrophysical systems, for example in solar and YSO flares \citep{somov1976physical,feigelson1999high}. In these systems, emission may even be strong enough to cause significant cooling of the plasma \citep{somov1976physical,oreshina1998slow}. Magnetic reconnection has also been postulated to be responsible for the high-energy radiation observed from many extreme relativistic astrophysical environments, such as black hole accretion disks and their coronae \citep{goodman2008reconnection,beloborodov2017radiative,werner2019particle,ripperda2020magnetic,mehlhaff2021pair,hakobyan2023radiative,chen2023synchrotron}, gamma-ray bursts (GRBs) \citep{lyutikov2006electromagnetic,giannios2008prompt,zhang2010internal,uzdensky2011magneticb,mckinney2012reconnection}, pulsar magnetospheres \citep{lyubarskii1996generation,lyubarsky2001reconnection,zenitani2001generation,zenitani2007particle,jaroschek2009radiation,uzdensky2013physical,cerutti2015particle,cerutti2016modelling,philippov2018ab,philippov2019pulsar,hakobyan2019effects,hakobyan2023magnetic}, pulsar wind nebulae \citep{uzdensky2011reconnection_powered,cerutti2012extreme,cerutti2013simulations,cerutti2017dissipation,cerutti2014three}, magnetar magnetospheres \citep{lyutikov2003explosive,uzdensky2011magneticb,schoeffler2019bright,schoeffler2023high}, and and in jets from active galactic nuclei (AGN) \citep{romanova1992magnetic,jaroschek2004relativistic,giannios2009fast,petropoulou2023hadronic,nalewajko2011radiative,nalewajko2014constraining,sironi2015relativistic,mehlhaff2020kinetic,mehlhaff2021pair}. In these extreme astrophysical systems, reconnection occurs in a regime where other radiative effects, such as Compton drag and radiation pressure, can further influence the reconnection process \citep{uzdensky2011magnetic,uzdensky2011magneticb,uzdensky2016radiative}.

In this paper, we focus on the effects of radiative cooling, which results in the rapid removal of internal energy from the reconnecting system. A discussion of other radiative effects is provided in  \cite{uzdensky2011magneticb, uzdensky2016radiative}. Dominant cooling mechanisms vary among astrophysical environments --- some examples include bremsstrahlung emission in the solar corona \citep{krucker2008hard}, line and recombination emission from ionization fronts in astrophysical jets \citep{blondin1989herbig,masciadri2001optically}, synchrotron cooling in pulsar magnetospheres, pulsar wind nebulae, and magnetar magnetospheres \citep{lyubarsky2001reconnection,uzdensky2013physical,cerutti2015particle,cerutti2016modelling,uzdensky2011reconnection_powered,schoeffler2023high,chernoglazov2023high}, and inverse-Compton cooling in black hole coronae \citep{goodman2008reconnection,beloborodov2017radiative,werner2019particle,sironi2020kinetic,sridhar2021comptonization}. Radiative cooling becomes important when the radiative cooling time of a fluid element becomes comparable to the time spent inside the reconnection layer (also called the current sheet) \citep{uzdensky2016radiative}. We can quantify the importance of radiative cooling using the dimensionless cooling parameter $R \equiv \tau_{\text{cool}}^{-1} / \tau_A^{-1}$, which describes radiative cooling rate $\tau_{\text{cool}}^{-1} = P_{\text{rad}}/E_{\text{th}}$ relative to the Alfvénic transit rate $\tau_A^{-1} = V_A/L$. Here, $E_{\text{th}} = p_{\text{th}}/(\gamma -1)$ is the thermal energy density which depends on the pressure $p_{\text{th}}$ and the adiabatic index $\gamma$, $P_{\text{rad}}$ is the volumetric radiative power loss, and $L$ is the size of the reconnection layer. When $R_\text{cool} \gtrsim 1$, reconnection occurs in the radiatively-cooled regime.

\cite{uzdensky2011magnetic}, building upon earlier work by \cite{dorman1995one}, provided the first theoretical description of reconnection in radiatively-cooled collisional plasmas.  Allowing for radiative losses and compressibility in the classical Sweet-Parker theory \citep{parker1957sweet}, they predicted three primary effects of radiative cooling --- (1) radiative cooling limits the temperature rise of the reconnection layer, generating a colder layer compared to the non-radiative case; (2) there is strong compression of the reconnection layer, generating a denser thinner layer; and (3)  radiative cooling instabilities can generate rapidly growing perturbations that disrupt the current sheet \citep{uzdensky2011magnetic, uzdensky2011magneticb, uzdensky2016radiative}. The colder layer temperature is a consequence of energy balance within the reconnection layer, since Ohmic heating must also balance radiative losses in addition to the enthalpy leaving the layer in the outflows. 
Since the plasma (Spitzer) resistivity scales with temperature as $\bar{\eta} \sim T^{-3/2}$ \citep{chen1984introduction}, a lower temperature leads to a more resistive layer, and the Lundquist number $S_L = V_A L / \bar{\eta}$ becomes smaller. In the compressible Sweet-Parker model, the reconnection rate $E/B_\text{in}V_\text{A} \sim A^{1/2} S_L^{-1/2}$ also depends on the density compression ratio $A \equiv \rho_\text{layer}/\rho_\text{in}$  \citep{uzdensky2011magnetic}. The strong-compression solution $A \gg 1$ depends on the functional form of the dominant radiative loss mechanism $P_\text{rad}$. Strong compression $A \gg 1$ occurs for the case where Ohmic dissipation $\dot{q}_\text{Ohm} \approx A (B_\text{in}^2/\mu_0) (V_\text{A,in}/L)$ is primarily balanced by radiative losses $\dot{q}_\text{Ohm} \approx \dot{q}_\text{rad} $ \citep{uzdensky2011magnetic}. The combined effect of strong compression and the smaller Lundquist number results in faster reconnection rates in the radiatively-cooled regime. 

In the strongly radiatively-cooled regime, the reconnection layer may be susceptible to radiative cooling instabilities. One such instability is the radiative collapse of the layer, which occurs when cooling induces dynamics that further increase the cooling rate, and result in ever-increasing compression of the layer \citep{dorman1995one, uzdensky2011magnetic}. The layer is unstable to radiative collapse if the function $P_\text{rad}(A)/\dot{q}_\text{Ohm}(A)$ has a positive derivative with respect to $A$, i.e. an increase in compression of the layer causes radiative losses to increase faster than Ohmic dissipation, in turn leading to more compression. In addition to radiative collapse, the reconnection layer may also be susceptible to a host of thermal-condensation instabilities, and the coupling of these thermal instabilities with the tearing instability can be important for the transient dynamics of the reconnection process \citep{somov1976physical,steinolfson1984radiative,tachi1985radiative,forbes1991numerical,oreshina1998slow,jaroschek2009radiation,sen2022}. 

Although radiative cooling is important in many astrophysical plasmas, radiatively-cooled magnetic reconnection is not adequately understood, which has motivated several numerical studies of radiative reconnection \citep{forbes1991numerical,oreshina1998slow,jaroschek2009radiation,laguna2017effect,ni2018magnetic}. These studies are consistent with the predictions of \cite{uzdensky2011magnetic}, showing denser, thinner, and colder current sheets with faster reconnection rates \citep{oreshina1998slow,laguna2017effect,ni2018magnetic,ni2018magnetic2}. Furthermore, these simulations also show decreased outflow velocity in the radiatively-cooled case, since part of the dissipated magnetic energy is lost via radiative emission from the layer \citep{oreshina1998slow}. Numerical studies also show evidence for run-away compression of the layer \citep{dorman1995one, schoeffler2019bright, schoeffler2023high}, and for the onset of thermal-condensation instabilities \citep{forbes1991numerical,oreshina1998slow}. In recent years, there has also been an explosion in the number of radiative-PIC (particle-in-cell) simulations of (relativistic) magnetic reconnection, for the modeling of reconnection physics in extreme astrophysical systems \citep{jaroschek2009radiation, cerutti2014three,cerutti2013simulations, werner2019particle, mehlhaff2020kinetic,mehlhaff2021pair, schoeffler2019bright, schoeffler2023high, sironi2020kinetic, sridhar2021comptonization, sridhar2023comptonization, hakobyan2019effects, hakobyan2023radiative,chernoglazov2023high}. Radiative-PIC simulations of current sheets unstable to the plasmoid instability in electron-positron pair plasmas have shown strong cooling-driven compression of the density and reconnected magnetic flux inside the plasmoids, making them sites of enhanced radiative emission \citep{schoeffler2019bright,schoeffler2023high}.


\begin{figure}
\includegraphics[page=8,width=1.0\textwidth]{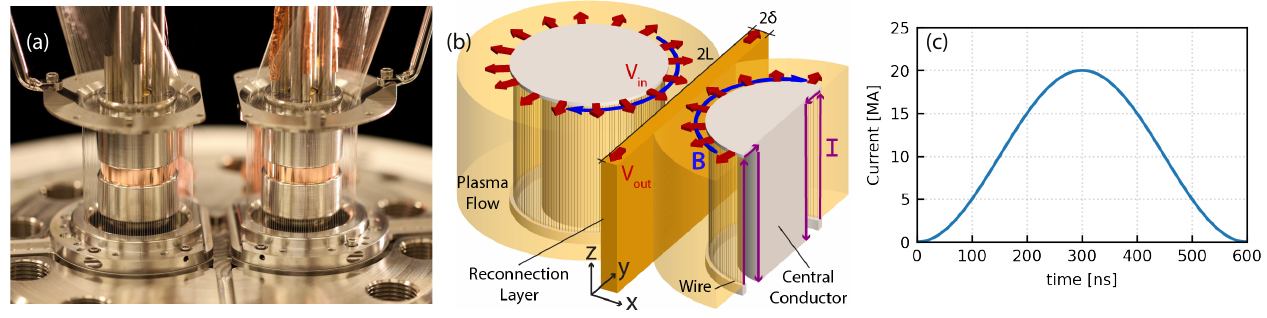}
\centering
\caption{(a) A photograph of the experimental load hardware for the MARZ experiments, which is the geometry we simulate using GORGON. Each array consists of 150 Al wires, \SI{75}{\um} diameter, evenly spaced in a cylinder \SI{40}{\mm} in diameter, \SI{40}{\mm} tall, with a center-to-center separation of \SI{60}{\mm}, or \SI{20}{\mm} between the wires of opposing arrays. (b) A three-dimensional model of the load hardware showing the direction of the current flow (purple), plasma ablation (red), and advected magnetic field (blue). The flows from each array interact in the mid-plane to form a current sheet.
(c) The current pulse used in the Z experiments and the GORGON simulations, which is well approximated by $I = I_0 \sin^2(\pi t/2\tau)$ with $I_0 = \SI{20}{\mega\ampere}$ and $\tau = \SI{300}{\ns}$.}
\label{fig:load}
\end{figure}

\subsection{Radiatively-cooled reconnection in the laboratory}

Despite the promising results of numerical simulations, there have been few experimental studies of radiatively-cooled reconnection in the laboratory. The primary reason for this is the difficulty associated with achieving the plasma conditions required for observing radiative cooling effects on experimental time scales. As an example, \autoref{tab:table1} summarizes the working conditions of some major reconnection experiments. We calculate the cooling time using an optically thin radiative loss model for simplicity, although more sophisticated radiation loss models which account for opacity and non-equilibrium emission can also be used for this calculation \citep{hare2018experimental}. For MRX \citep{ji1999magnetic,yamada2010magnetic} and laser-driven experiments \citep{rosenberg2015slowing,fox2011fast,fox2012magnetic}, which have fully stripped ions, we use a recombination-bremsstrahlung model \citep{richardsonnrl2019}, whereas, for the pulsed-power-driven experiments \citep{suttle2016structure,suttle2018ion,hare2018experimental}, we use emissivities calculated with the atomic code SpK, which includes line, bremsstrahlung, and recombination emission \citep{crilly2022spk}. Inverse-Compton and cyclotron/synchrotron radiation mechanisms are not included, and are not expected to be significant. Of the experiments listed in  \autoref{tab:table1}, pulsed-power-driven reconnection experiments exhibit the largest cooling parameter. Indeed, previous pulsed-power driven experiments on 1 MA university-scale facilities have provided evidence for the onset of radiative cooling ---  Thompson scattering data show strong cooling of the ions in the reconnection layer \citep{suttle2018ion, hare2018experimental}. However, these pulsed-power-driven experiments conducted on 1 MA machines either achieve strong cooling at low ($<10$) Lundquist numbers \citep{suttle2016structure, suttle2018ion, hare2018experimental}, or little cooling at relatively higher ($\sim 100$) Lundquist numbers \citep{hare2017formation,hare2017anomalous,hare2018experimental}. In contrast, the pulsed-power experiments simulated here will simultaneously achieve both a higher Lundquist number and a high cooling parameter.

\begin{table*}
    \centering
    \ra{1.3}
    \begin{tabular}{l  c c c c}
    \hline
         & MRX & Laser-driven  & Pulsed-power (1 MA) &  MARZ \\
         \hline 
        $S_L$  & $>500$ & 100 & 10 & 400 \\
        $n_e$ ($\SI{}{\per \cubic \centi \metre}$) & $5 \times 10^{13}$ & $1 \times 10^{20}$ & $1 \times 10^{18}$ & $1 \times 10^{20}$ \\
        $T_e$ (\SI{}{\electronvolt}) & 10 & 1300 & 40 & 100 \\
        B (T) & 0.1 & 55 & 3 & 50 \\
        $V_A$ ($\SI{}{\kilo \metre \per \second})$ & 50 & 65 & 30 & 70 \\
        \hline
        $\tau_\text{cool}$ (ns) & $1.2 \times 10^9$ & $1300$ & 0.2\footnotemark & 1 \\
        $\tau_A$ (ns) & $1.6 \times 10^4$ & 2 & 230 & 210 \\
        $R_\text{cool}$ & $1 \times 10^{-5}$ & 0.03 & 1000 & 240 \\
        \hline
    \end{tabular}
    \caption{Comparison of characteristic working conditions in laboratory experiments of magnetic reconnection, between MRX \citep{ji1999magnetic,yamada2010magnetic}, laser-driven reconnection \citep{rosenberg2015slowing}, 1 MA pulsed-power \citep{suttle2018ion,hare2018experimental}, and the MARZ experiments. Quantities above the horizontal line are reported values, while quantities below the line are calculated from the reported values using optically-thin radiation models. For MRX and laser-driven experiments, which have fully stripped ions, we use a recombination-bremsstrahlung model \citep{richardsonnrl2019}, whereas, for the pulsed-power-driven experiments, we use emissivities calculated with the atomic code SpK \citep{crilly2022spk}.}
     \label{tab:table1}
\end{table*}

\footnotetext{\cite{suttle2018ion} report a cooling time of 5 ns. However, using SpK as discussed below results in a shorter cooling time for relevant densities and temperatures.}

The simulations presented in this paper were motivated by experiments run by the Magnetic Reconnection on Z (MARZ) collaboration, which uses the Z Machine at Sandia National Laboratories to investigate radiatively-cooled magnetic reconnection.
The Z machine is a pulsed-power generator that delivers peak currents of $20-\SI{30}{\mega\ampere}$ with $100-\SI{300}{\ns}$ rise times to a load inside a vacuum chamber  \citep{sinars2020review}.
For the MARZ experiments, we scale up the pulsed-power-driven magnetic reconnection platform developed on the MAGPIE generator at Imperial College London \citep{suttle2016structure,  suttle2018ion, hare2018experimental}, which consists of two inverse or ``exploding'' cylindrical wire arrays, placed side by side and driven in parallel.  \autoref{fig:load}a shows a photograph of the load for the first MARZ experiment. Each array consists of 150 aluminum wires, $\SI{75}{\um}$ diameter arranged in a cylinder \SI{40}{\mm} in diameter around a thick central conductor.

The current from the generator passes through the wires and returns to ground through the central conductor, Ohmically heating the wires until they undergo an ``electrical explosion'', and form a heterogeneous liquid-droplet/vapor mixture. Further Ohmic heating forms a coronal plasma around each wire, which is accelerated radially outwards by the $\textbf{j}\times \textbf{B}$ force due to the strong azimuthal magnetic field ($\approx 100$ T) around the central conductor. 
As the plasma moves away from the wire, it advects with it some of this driving field, creating radially-diverging supersonic super-Alfvénic outflows with frozen-in magnetic fields \citep{burdiak2017structure,suttle2019interactions,datta2022structure,datta2022time}. This process is referred to as ablation, and in the MARZ experiments, we choose an initial wire diameter such that the arrays are ``over-massed'', and the wires act as stationary reservoirs of mass throughout the current pulse \citep{harvey2009quantitative,lebedev2001effect,datta2023plasma}. 

When the radially accelerated plasma flows from the two wire arrays collide at the mid-plane, the advected magnetic fields are equal in magnitude and anti-parallel (see \autoref{fig:load}b). A current sheet forms at the mid-plane, and magnetic reconnection occurs. In these experiments, the plasma cools through a combination of bremsstrahlung, recombination, and line emission during the reconnection process. The cooling mechanisms in these laboratory experiments are therefore not the same as those in the extreme astrophysical plasmas discussed above, where synchrotron and other mechanisms are often more important.  Although this is a limitation of these experiments, we are still qualitatively in the same regime, in which radiative cooling timescales are short enough to affect the dynamics of magnetic reconnection. As seen in \autoref{tab:table1}, pulsed-power-driven reconnection experiments achieve cooling parameters several orders of magnitude higher than other types of reconnection experiments.

In these experiments, the plasma flows are highly collisional ($\lambda_{ii} \sim 0.1-\qty{1e-2}{\milli \meter}$), and therefore, well approximated by magnetohydrodynamics (MHD) \citep{suttle2019interactions}. The inflows to the reconnection layer are axially uniform, so any three-dimensional dynamics within the layer are the result of instabilities rather than the inflows. The driving current pulse is much longer than the Alfv\'en transit time so the inflows can be considered in approximate steady-state, and rapid changes in the plasma dynamics are again the result of instabilities rather than the changing drive conditions.
As we simulate the entire experimental domain from the start of the current pulse, we are inherently simulating a forming current sheet, rather than starting with an initial condition such as a Harris sheet.

In this paper, we present two-dimensional and three-dimensional MHD simulations of the MARZ experiments. To elucidate the effects of radiative cooling, we compare our 2D results for the radiatively-cooled and non-radiative cases, with a non-optically thin radiative loss model computed using the atomic code SpK \citep{crilly2022spk}. In both the non-radiative and radiatively-cooled cases, the arrays generate magnetized supersonic ($M_S = 4-5$), super-Alfvénic ($M_A \approx 1.5$), and super-fast magnetosonic ($M_{FMS} \approx 1.4$) flows which interact in the mid-plane to generate a current sheet. The current sheet exhibits a heterogeneous structure due to the presence of several fast-moving plasmoids. These plasmoids are sites of strong radiative emission due to their higher density and temperature compared to the rest of the layer, similar to observations in previous numerical studies of radiative reconnection \citep{sironi2020kinetic, schoeffler2019bright,schoeffler2023high}. We find that radiative cooling modifies the reconnection process in several ways. First, it creates a denser, colder, and thinner reconnection layer that exhibits strong compression, consistent with the theoretical prediction of \cite{uzdensky2011magnetic}. Second, the current sheet also becomes more uniform due to the cooling-driven extinction of plasmoids in the current sheet. Finally, there is also reduced flux pile-up outside the layer, resulting in a lower magnetic field and density of the inflows into the sheet. The dynamics observed in the two-dimensional simulations are well reproduced in 3D. Furthermore, the plasmoids in the 3D simulation also exhibit strong kinking along the axial direction. Radiation transport significantly modifies the inflow into the current sheet in both 2D and 3D, resulting in an initial lower driving magnetic pressure, which in turn, causes reduced compression of the layer after radiative collapse.

\section{Simulation Setup}
\label{sec:setup}

We perform compressible resistive-MHD simulations of a dual exploding wire array load using the code GORGON. GORGON is a three-dimensional (cartesian, cylindrical, or polar coordinates) Eulerian resistive-MHD code with van Leer advection \citep{Chittenden2004, Ciardi2007}. The simulation geometry consists of two exploding wire arrays with a center-to-center separation of $\SI{60}{\milli \meter}$. Each array has a diameter of $\SI{40}{\milli \meter}$, and consists of 150 equally-spaced $\SI{75}{\micro \meter}$ diameter aluminum wires. In 3D, the wires are \SI{36}{\milli\meter} tall. The wire arrays are over-massed to provide continuous plasma ablation without exploding during the simulation. The initial mass in the wires is distributed over $3 \times 3$ grid cells of pre-expanded wire cores. The current is applied to the wire array by setting the magnetic field in the region between the central conductor and the wires, using a current pulse of the form $I = I_0 \sin^2(\pi t/2\tau)$ with $I_0 = \SI{20}{\mega\ampere}$ and $\tau = \SI{300}{\ns}$ (\autoref{fig:load}c),  chosen to simulate the Z machine's current pulse when operated in long-pulse mode \citep{sinars2020review}.
 
 We first perform two-dimensional simulations in the $xy-$plane (see \autoref{fig:load}b) on a $3200 \times 1760$ cartesian grid of dimensions $160 \times 88$ \SI{}{\milli\metre\squared}. The grid cell size is $\Delta x = \SI{50}{\micro \meter}$, which is adequate to resolve the resistive diffusion length $\bar{\eta}/V > 4 \Delta x$, calculated using the magnetic diffusivity $\bar{\eta}$ of the reconnection layer, and the inflow velocity $V$. Two-fluid effects are not included in these simulations, and only the resistive-MHD equations are solved. 
 Open boundary conditions are imposed on all sides of the computational domain. GORGON uses an adaptive time-step, and we output the results of the simulation every \SI{10}{\nano \second}. The 2D simulations were run for $2 \tau = \SI{600}{\nano \second}$, which is roughly $300$ times the Alfvén crossing time $\delta/V_A$. Here, we have used averaged values of the Alfvén speed $V_A = B_{in}/\sqrt{\mu_0 \rho_{in}} \approx \SI{50}{\kilo\meter\per\second}$, calculated just outside the reconnection layer, and the reconnection layer half-width $\delta \approx \SI{0.1}{\milli\meter}$ at the time of peak current in the radiatively-cooled simulation. 
 
 Three-dimensional simulations were also performed by extending the simulation domain by $\qty{36}{\milli \meter}$ (720 grid cells) in the $z$ direction. The grid cell size is the same as that in the 2D simulations. Reflective boundary conditions are used on the top and bottom surfaces of the simulation domain, while the sides of the simulation have open boundary conditions. The 3D simulations, which are computationally more expensive, were run for $\SI{280}{\nano \second}$, adequate to observe the formation and radiative collapse of the reconnection layer. 

GORGON solves two coupled energy equations for the ions and electrons. Both the ions and electrons transport heat via thermal conduction, and are heated or cooled by compression or expansion.
The ions are additionally heated by viscous heating, while the electrons are heated by Ohmic dissipation. The ion and electron temperatures equilibrate at a collisional energy equilibration rate $\tau_{E}^{-1} =  \SI{3.2e-9}{}n_i\bar{Z}^2 \text{ln} \, \Lambda / (A T_e^{3/2})$, where $\bar{Z}$ is the ionization, $n_i$ and $T_e$ are the ion density and electron temperature respectively, $\text{ln} \, \Lambda$ is the Coulomb logarithm, and $A$ is the ion mass in proton mass units \citep{ciardi2007evolution,richardson2019}. The equilibration time is initially on the order of the Alfvén transit time $\tau_A = L/V_A \sim 4 \tau_E$, but becomes much shorter later in time ($\tau_{E}/\tau_A \sim \SI{e-4}{}$), such that the ion and electron temperatures become equal. Here, we calculate the Alfvén transit time using the Alfvén speed in the inflow to the reconnection layer, and L is the layer half-length $L \approx \SI{18}{\milli\meter}$ (see \S \ref{sec:results} for details on how these quantities are calculated). 
We use a Thomas-Fermi equation of state to determine the (isotropic) pressure and ionization level of the plasma \citep{Ciardi2007}. Transport coefficients are determined from \cite{epperlein1986plasma}, and vary spatially and temporally with changes in the plasma's electron temperature, density, average ionization, and the magnitude and orientation of the magnetic field. The electrons also lose internal energy via radiative losses --- accurate modeling of radiation is of particular importance in the description of the radiative collapse of the reconnection layer.

\subsection{Radiation models}\label{section:radmodels}

In the limit of negligible optical depth, radiation can be treated as an electron energy loss mechanism determined entirely by the plasma's total emissivity, $J$. As optical depth increases, radiation transport effects become increasingly important as radiation emitted in one region can be reabsorbed in another. Using an optically thin radiation model in plasmas with finite opacity would result in an overestimation of the total energy loss from the system. However, solving radiation transport in large MHD simulations is a computationally intensive task. Therefore, to limit the total radiative loss compared to optically thin models, we explore a local loss model in our simulations, which is computationally less expensive than solving radiation transport. The effects of the local loss model are compared to P$_{1/3}$ multi-group radiation transport. 

In the local loss model, the optical depth of the computational cell itself is included in the calculation of the radiative power emitted by each individual cell. For an isotropically-emitting spherical volume, an analytic solution for the radiative loss rate per unit volume, $P_{\text{rad}}$, can be found from the time-independent, frequency-resolved, radiation transport equation \citep{crilly2020simulation}:

\begin{align}\label{eqn:rad_tran}
    P_{\text{rad}} &= \frac{3}{4R} \int \frac{4\pi j_\nu}{\kappa_\nu}\left[1+\frac{2}{\tau_\nu^2}\left((1+\tau_\nu)e^{-\tau_\nu}-1\right)\right] dh\nu \ , \ \mbox{where} \ \tau_\nu = 2\kappa_\nu R \ ,  \\
    \lim_{\tau_\nu \rightarrow \infty} P_{\text{rad}} &= \frac{3}{4R} \int \frac{4\pi j_\nu}{\kappa_\nu} dh\nu = \frac{S \sigma T^4}{V} \ , \\
    \lim_{\tau_\nu \rightarrow 0} P_{\text{rad}} &= \int 4\pi j_\nu \ dh\nu = J \ ,
\end{align}
where $R$, $S$, and $V$ are the radius, surface area, and volume of the sphere, $j_\nu$ is the emissivity, and $\kappa_\nu$ is the absorption opacity. Scattering effects are not included in this model. For non-spherical volumes, such as the cubic computational cell used in these simulations, the radius is exchanged for the effective width of the cell as calculated by 3 times the volume-to-surface area ratio. As opposed to optically thin models, the optical depth of the computational cell limits the total radiative power lost from the cell. Radiation emitted by a given cell is, however, not re-absorbed by neighboring cells in the local loss model, and lost from the system. While this approximation neglects re-absorption over length scales longer than a computational cell, and thus still over-estimates the total radiative loss, it serves as an improvement over optically thin models as energy is retained by the system due to local re-absorption which would have otherwise been lost.

Numerically, a multi-group approach can be used to evaluate the local loss model for $P_{\text{rad}}$ using opacities and emissivities from tables. In GORGON, multi-group tables from the code SpK are used \citep{crilly2022spk}. SpK performs detailed configuration accounting calculations of electronic and ionic stage populations in either Local Thermal Equilibrium (LTE) or NLTE through an effective temperature approach. The radiation model includes free-free, free-bound, and bound-bound transitions from which opacities and emissivities are calculated, which are functions of the local ion density and electron temperature. 

\begin{figure} 
\includegraphics[page=1,width=1.0\textwidth]{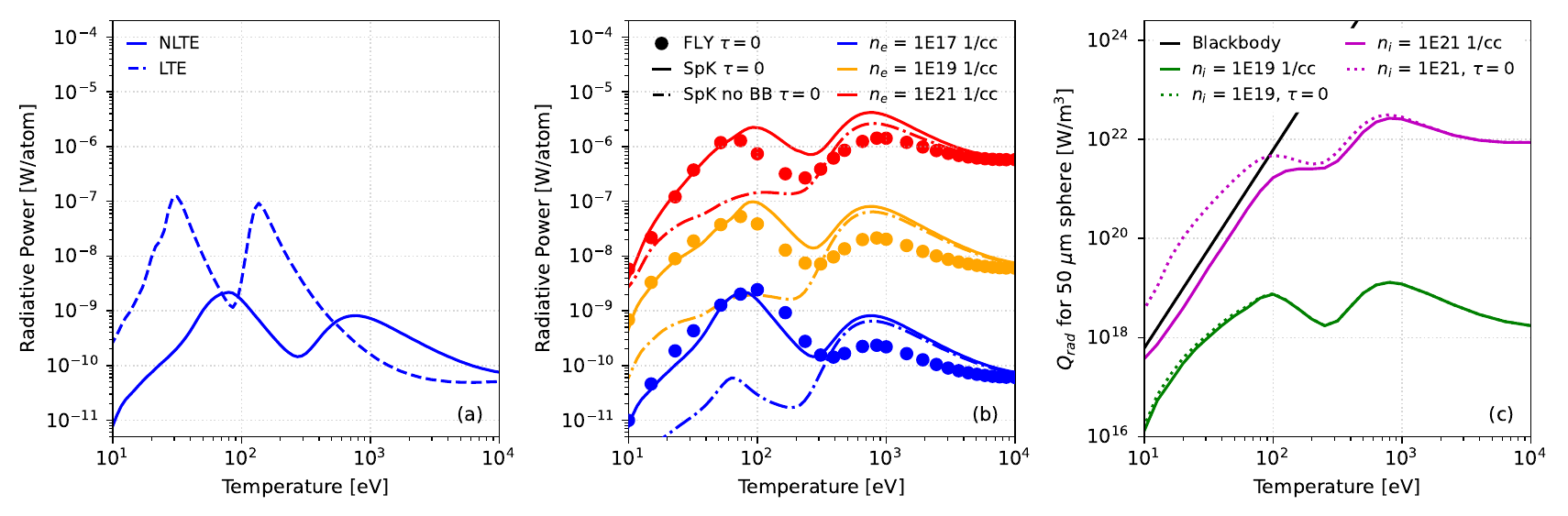}
\centering
\caption{(a) The optically thin ($\tau = 0$) radiative power of Al plasma at an electron density of 1$\times 10^{17}$ cm$^{-3}$ and varied temperature calculated by SpK in LTE and NLTE. (b) The optically thin radiative power at various electron densities and temperatures. Results from NLTE SpK with and without bound-bound (BB) transitions are compared to NLTE FLY tables \citep{Chung2003,Chung2005}. (c) SpK predictions of the local loss radiative power for a $R$ = 50 $\mu$m sphere as a function of temperature.}
\label{fig:radiativepower}
\end{figure}

Local loss models provide sufficient accuracy to perform design calculations and investigate the physical phenomena for the radiatively-cooled reconnection platform. \autoref{fig:radiativepower} shows results of the local loss model (\autoref{eqn:rad_tran}) for an Aluminium plasma. It is clear from \autoref{fig:radiativepower}a that an NLTE description is required to accurately calculate the radiative power at lower densities. The simulations in this paper, therefore, use NLTE opacity and emissivity tables from SpK, which are valid for the range in density and temperature accessible by the MARZ experiments. We note that a corresponding NLTE effect on the equation of state will exist but the corrections are considerably smaller than on the radiative power. It is also shown in \autoref{fig:radiativepower}b that L-shell line emission is dominant at temperatures around \SI{100}{\electronvolt}, thus continuum loss models which only include bremsstrahlung and radiative recombination are inaccurate. Additionally, the local loss model predicts large corrections to the radiative loss in denser plasma due to optical depth effects, as seen in \autoref{fig:radiativepower}c. 

For a more accurate description of the experiment, a limited number of 2D and 3D simulations were also run with P$_{1/3}$ multi-group radiation transport, the numerical implementation of which can be found in \citep{crilly2022spk}. In the P$_{1/3}$ multi-group radiation transport model, radiation emitted by a given cell can be absorbed and re-emitted by plasma in other parts of the simulation domain. This task, however, is computationally more expensive than the local loss model described above. We discuss the importance of radiation transport modeling in \S \ref{sec:transport_effect}.


\section{Two-dimensional simulations}

\subsection{Results}
\label{sec:results}

We describe and compare the 2D ($xy-$ plane) simulation results for two cases, first for the non-radiative case in which we artificially turn off all radiative losses from the plasma, and next for the radiatively-cooled case where the losses are implemented using the local loss radiation model described above.

\subsubsection{Non-radiative Case}
\label{sec:nonradiative}

\begin{figure} 
\includegraphics[page=1,width=1.0\textwidth]{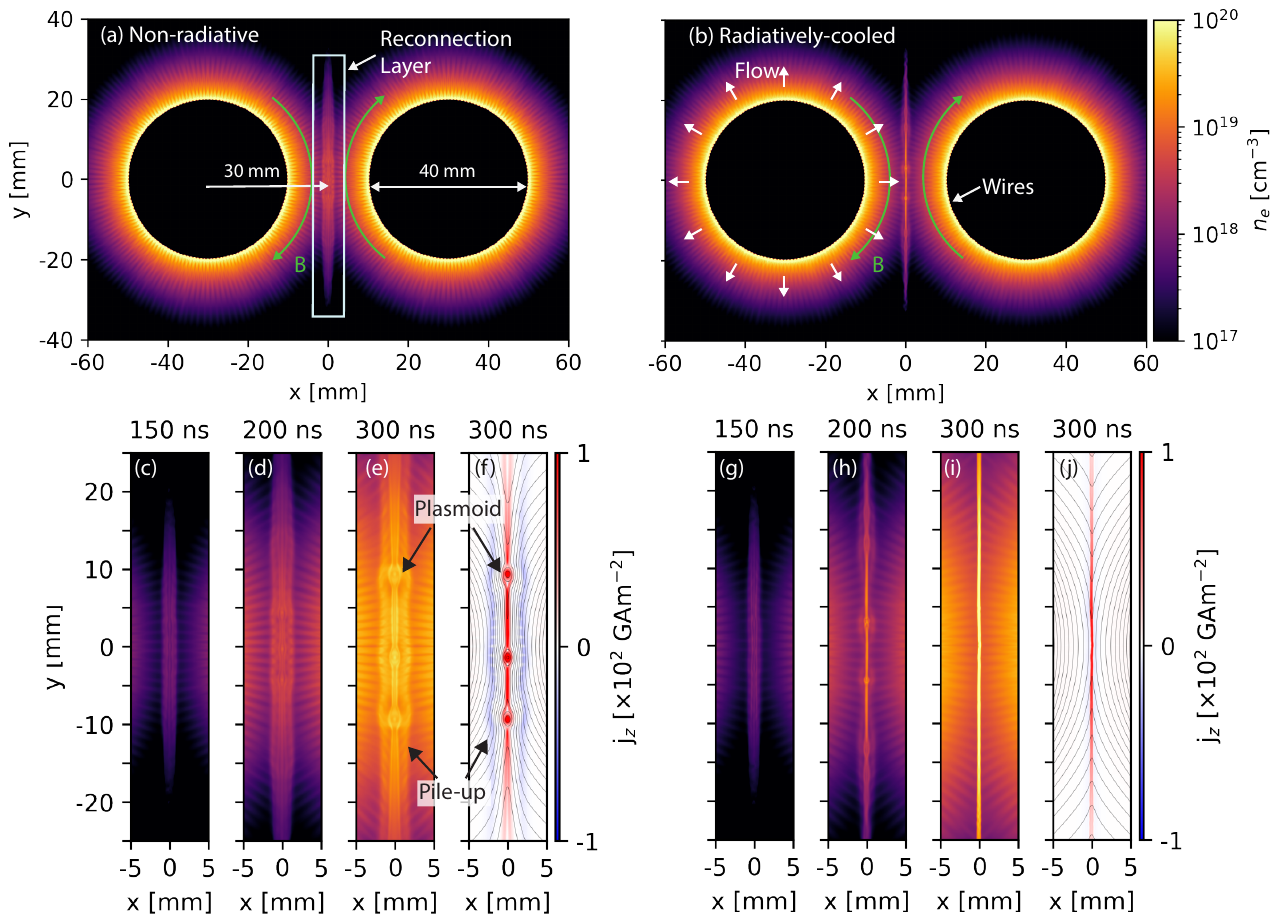}
\centering
\caption{ (a) Electron density at 200 ns after current start from two-dimensional resistive-MHD simulations for the non-radiative case. The wire arrays generate radially-diverging flows which interact at the mid-plane to generate a current sheet.
(b) Electron density at 200 ns after current start for the radiatively-cooled case. (c-d) Electron density in the reconnection layer at 150, 200, and 300 ns after current start for the non-radiative case showing the formation of plasmoids. (f) Current density and superimposed magnetic field lines in the reconnection layer at 300 ns after current start for the non-radiative case showing flux pile-ups and plasmoids. (g-i) Electron density in the reconnection layer at 150, 200, and 300 ns after current start for the radiatively-cooled case. (f) Current density and superimposed magnetic field lines in the reconnection layer at 300 ns after current start for the radiatively-cooled case. }
\label{fig:density}
\end{figure}

\autoref{fig:density}a shows the electron density distribution at $t = 200$ ns after the start of the current pulse for the case with no radiative emission. Each wire array generates radially-diverging plasma flows, so the electron density is high close to the wires and decreases with distance from the arrays. The electron density from each array also exhibits a periodic small-scale modulation in the azimuthal direction, due to the supersonic collision of adjacent azimuthally-expanding ablation flows from the individual wire cores \citep{Swadling2013}. This results in the formation of standing oblique shocks, periodically distributed in the azimuthal direction. The length scale of this azimuthal modulation is comparable to the inter-wire separation of around $\SI{0.8}{\milli \meter}$.

The plasma flows advect magnetic field from the inside of the array as they propagate radially outwards. The magnetic field lines are oriented azimuthally with respect to the center of each array. The plasma flows with oppositely-directed and symmetrically-driven magnetic fields interact at the mid-plane ($x = 0$) to generate a current sheet. The structure and time-evolution of the current sheet are shown in {\color{blue} Figures} \ref{fig:density}(c-f). The current sheet appears as an elongated region of enhanced current (see \autoref{fig:density}f) and electron density at the mid-plane. Magnetic field lines oriented in the $\pm y$-direction are driven into the current sheet by the inflows, and exit the reconnection layer as curved reconnected field lines, as seen in \autoref{fig:density}f. The current sheet first forms at $t \approx \SI{100}{\nano \second}$, consistent with the transit time between the wire locations and the mid-plane, and a flow velocity of $\SI{100}{\kilo \meter \per \second}$ \citep{hare2017formation,suttle2018ion}.  The current and electron density in the sheet increase with time. This is due to increased ablation from the wires as the magnitude of the driving current ramps up over time.

\begin{figure}
\includegraphics[page=3,width=1.0\textwidth]{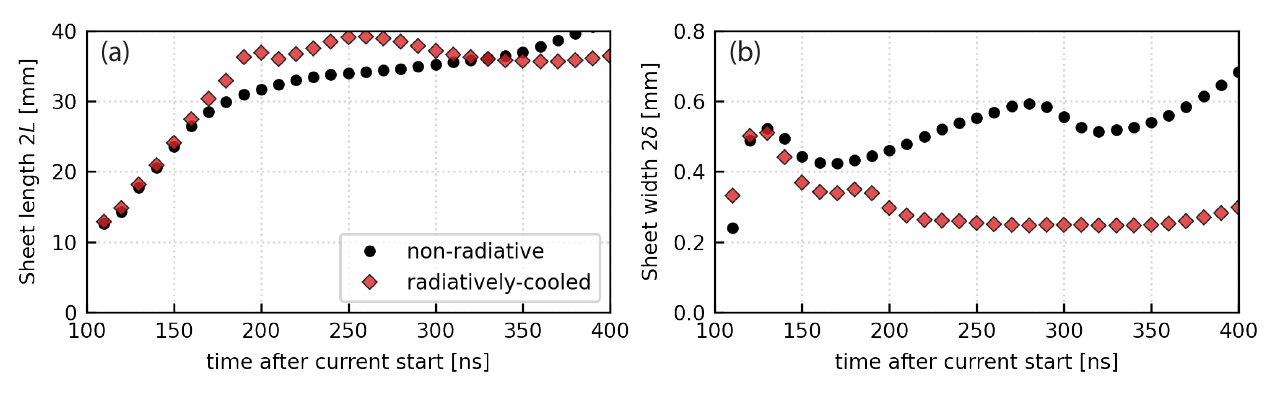}
\centering
\caption{ Variation of current sheet length (a) and current sheet width (b) with time for the non-radiative and radiatively-cooled cases.}
\label{fig:size}
\end{figure}

\autoref{fig:size} shows the temporal evolution of the length $2L$ and width $2 \delta$ of the current sheet. We define $2L$ as the full width at half maximum (FWHM) of the out-of-plane current density $j_z$ in the $y-$direction. To calculate the length of the current sheet, we first integrate $j_z$ in the $x$-direction between $-\SI{1}{\milli \meter} \leq x \leq \SI{1}{\milli \meter}$, then compute the FWHM of a Gaussian fit to the line-integrated current density. Similarly, to calculate the sheet width, we first integrate $j_z$ in $y$ between $-L \leq y \leq L$. We define the sheet width $2\delta$ based on a Harris sheet profile $\left[B_y(x) = \text{tanh}(x/\delta); j_z = \text{sech}^2(x/\delta)/\delta\right]$. For a Harris sheet, $j_z$ falls to 10\% of its peak value at $x=x_{10}=\pm1.82\delta$, so $\delta$ can be calculated as $\delta \approx x_{10}/1.82$. For Harris-like current sheets, $\delta$ estimated via the aforementioned method will be consistent with that approximated from the FWHM of $j_z$, i.e. $2\delta \approx \text{FWHM}/ 0.9$. In our simulations, $j_z$ appears Harris-like for the non-radiative case, but becomes flat-topped for the radiatively-cooled case. Using the FWHM to estimate $\delta$ in the radiatively-cooled case results in an overestimate of the sheet width, while using $\delta \approx x_{10}/1.82$ provides results that more appropriately capture the current sheet width. We use $10\%$ of the peak $j_z$ for this calculation in order to capture most of the current distribution.


For the non-radiative case (black circles in \autoref{fig:size}), the sheet length initially increases rapidly with time ($t < \SI{200}{\nano\second}$), and then continues to rise at a much slower rate. After the early transient period, the value of $2L \approx \SI{35}{\milli\meter}$ is comparable to the radius of curvature of the field lines at the current sheet. The width of the current sheet also exhibits an increase with time; the increase in $2\delta$ is modest, and the sheet width remains between $\SI{0.4}{\mm} \leq 2\delta \leq \SI{0.6}{\milli \meter}$ during $150-\SI{350}{\nano\second}$. The aspect ratio of the sheet after the formation stage is thus $\delta/L \approx 0.01$. Both $2L$ and $2\delta$ also increase faster later in time ($t \geq \SI{350}{\nano \second}$). This is related to a change in the ablation conditions due to explosion of the wire array, as the wires begin to run out of mass at this late time. In this paper, however, we are interested in the reconnection dynamics well before this late time. 

The current sheet exhibits a non-uniform structure, with elliptical islands of higher electron density separated by thin elongated regions. These density concentrations correspond to the locations of magnetic islands or `plasmoids'. This can be observed in  \autoref{fig:density}f, which illustrates the distribution of current density $j_z$ with superimposed magnetic field lines. The presence of plasmoids is consistent with magnetic reconnection at the current sheet, and indicates that the current sheet is unstable to the plasmoid instability \citep{loureiro2007instability}. The plasmoids envelop magnetic O-points in the reconnection layer, and are separated by individual X-points. More discussion on the structure and temporal evolution of the plasmoids is provided in \S \ref{sec:plasmoids}. 

{\color{blue} Figures} \ref{fig:density}(c-f) also show the presence of shocks upstream of the current sheet. Each shock appears as a discontinuous enhancement of the electron density in {\color{blue} Figures} \ref{fig:density}(c-e), and a thin region of negative current density in \autoref{fig:density}f. The presence of the shocks upstream of the current sheet is consistent with magnetic flux pile-up in a compressible system with super-magnetosonic inflows. Magnetic flux pile-up is expected to occur when the flux injection rate exceeds the flux annihilation rate in the reconnection layer \citep{biskamp1986magnetic}. We discuss flux pile-up in more detail in \S \ref{sec:pileup}.

\begin{figure}
\centering
\includegraphics[page=2,width=1\textwidth]{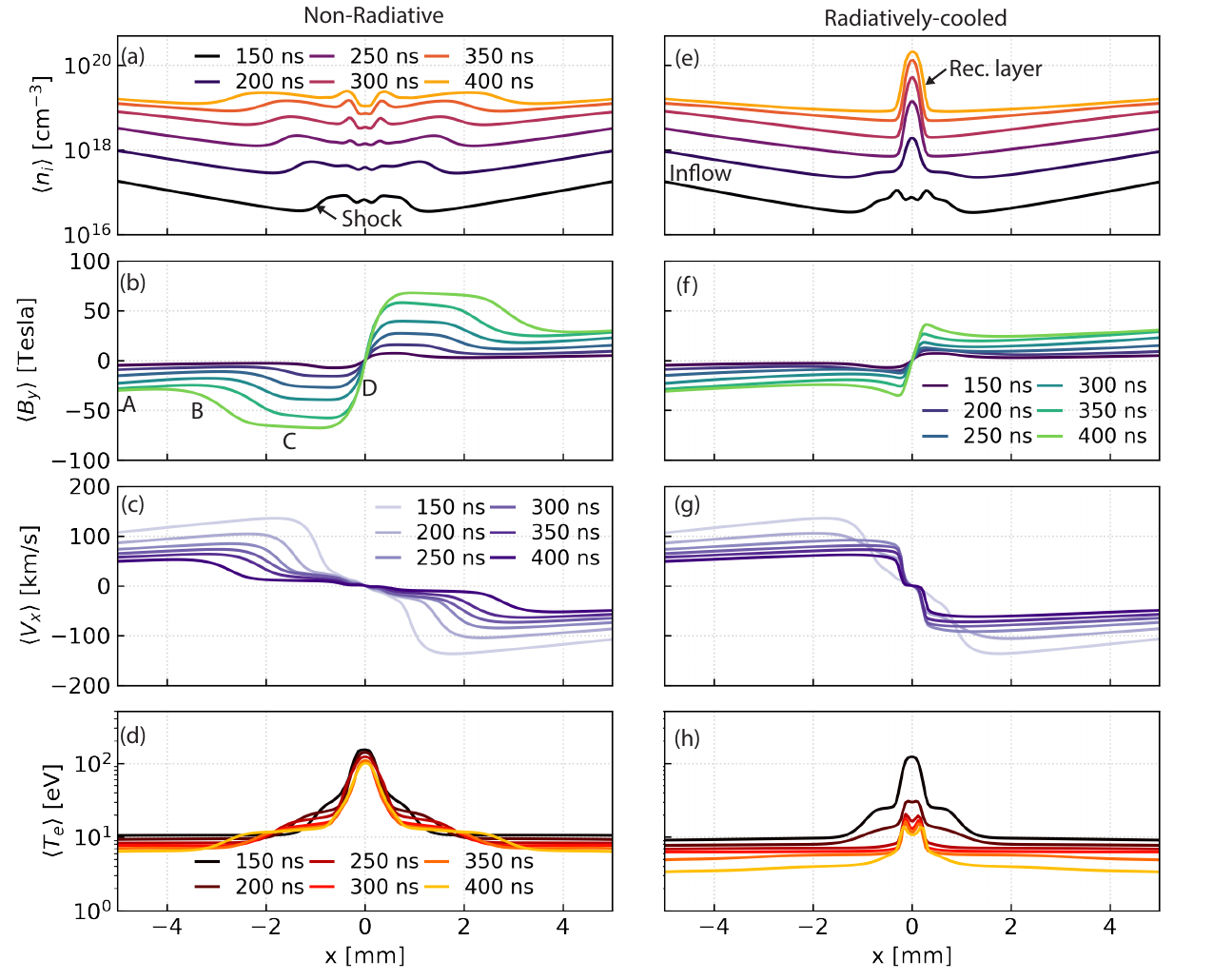}
\centering
\caption{Lineouts of ion density, magnetic field (y-component), flow velocity (x-component), and electron temperature as a function of $x$ for the non-radiative case (a-d), and the radiatively-cooled case (e-h). 
 In the non-radiative case, we see significant flux pile-up outside the layer, which leads to a $\approx 2 \times$ compression of the magnetic field and density in the inflows to the reconnection layer. In the radiatively-cooled case, we observe reduced flux pile-up and strong compression and cooling of the current sheet.}
 \label{fig:lineouts}
\end{figure}

{\color{blue} Figures} \ref{fig:lineouts}(a-d) show the lineouts of ion density $n_i$, the $y$-component of the magnetic field $B_y$, the $x$-component of the velocity field $V_x$, and the electron temperature $T_e$. The lineouts are taken along the $x$-axis, and each quantity is line-averaged in the $y$-direction between  $ -L/2 < y < L/2$. As shown in  \autoref{fig:lineouts}b, magnetic flux pile-up divides the plasma into 4 distinct regions --- (A) an inflow region upstream of the shock, (B) the shock transition region, (C) a post-shock region, and finally, (D) the reconnection layer. 

Consistent with time-of-flight effects and radially-diverging flow, the ion density and the magnetic field strength fall with increasing distance from the wires in the inflow region. The shock results in compression of both the ion density and the magnetic field by a factor of about $2$, while the velocity exhibits a sharp downward jump at the shock front. The sharp gradient in the magnetic field at the shock is consistent with the negative current density $\mu_0 j_z = \partial_x B_y - \partial_y B_x$ observed in  \autoref{fig:density}f, as expected from Ampere's law. The temperature also increases at the shock front due to compressional heating. The shocks propagate upstream with a velocity of about $\SI{10}{\kilo\meter\per\second}$, around $10\%$ of the inflow velocity.

The magnetic field continues to exhibit a gradual pile-up in the post-shock region, while the density decreases behind the propagating shock wave. As expected, the $y$-component of the magnetic field and the $x$-component of velocity undergo a reversal in direction inside the reconnection layer. The magnetic field $B_y$ and the inflow velocity $V_x$ approach 0 at the center of the reconnection layer ($x = \qty{0}{\milli \meter}$). The mass density inside the reconnection layer is similar to that just outside of the layer, indicating weak compression, while the electron temperature at the center of the layer is significantly higher ($T_e \approx \SI{100}{\electronvolt}$) than that just outside the layer ($T_e \approx \SI{10}{\electronvolt}$). This is consistent with the Ohmic dissipation of magnetic energy into internal energy during reconnection. Because of the temporal change in the driving current, the ion density and magnetic field increase with time, consistent with increased ablation from the wire arrays. The electron temperature, however, remains roughly constant with a value of $T_e \approx \SI{10}{\electronvolt}$ in the inflow, and $T_e \approx \SI{100}{\electronvolt}$ in the reconnection layer.

\subsubsection{Radiatively-Cooled Case}
\label{sec:radiative}

 \autoref{fig:density}b shows the electron density distribution from the wire arrays at $t = \SI{200}{\nano \second}$ for the radiatively-cooled case. Similarly, {\color{blue} Figures} \ref{fig:density}(g-j) show the electron density and current distribution in the reconnection layer for the radiatively-cooled case. The plasma outflows from the arrays, which are inflows into the reconnection layer, appear qualitatively similar to the non-radiative case. Early in time ($t < \SI{200}{\nano \second}$), the structure of the current sheet, and that of the upstream shock, is also similar to that in the non-radiative case. Lineouts along the $x$-axis [{\color{blue} Figures} \ref{fig:lineouts}(e-g)] show that the magnitudes of the line-averaged ion density, magnetic field $B_y$, and inflow velocity $V_x$ in the inflow region far from the current sheet remain almost identical to the non-radiative case. The electron temperature in the inflow is also similar to the non-radiative case early in time ($t = \SI{150}{\nano \second}$). However, as a consequence of radiative cooling, $T_e$ in the inflow (\SI{2.5}{\electronvolt} at \SI{400}{\nano\second}) becomes lower than the non-radiative inflow temperature (\SI{8}{\electronvolt} at \SI{400}{\nano\second}) later in time (\autoref{fig:lineouts}h).

The structure of the current sheet exhibits significant differences after $t \geq \SI{200}{\nano \second}$. {\color{blue} Figures} \ref{fig:density}(h-j) show a much thinner and denser current sheet than in the non-radiative case. In \autoref{fig:size}b, we compare the length $2L$ and width $2\delta$ of the current sheet with that for the non-radiative case. Initially, the dimensions of the current sheet in both cases are almost identical. For $t \geq \SI{200}{\nano \second}$, however, the radiatively-cooled current sheet becomes much thinner than in the non-radiative case, whereas the length remains approximately equal in the two cases. This results in a significantly smaller aspect ratio $\delta/L$ in the radiatively-cooled case. Moreover, whereas in the non-radiative case, we observe a modest increase in layer width over time, in the radiatively-cooled case, $2\delta$ is remarkably mostly constant within the interval $\SI{220}{\nano \second} \leq t \leq \SI{350}{\nano \second}$ (\autoref{fig:size}). 

The higher density and smaller width of the current sheet indicate strong compression of the current sheet due to radiative cooling. This can also be observed in lineouts of the ion density along the $x$-axis (\autoref{fig:lineouts}e), which show significantly higher density in the reconnection layer after $t = \qty{200}{\nano \second}$. The strong compression in the layer is indicative of radiative collapse. Evidence of radiative collapse is also observed from the significant decrease in the temperature in the layer (\autoref{fig:lineouts}h), which falls from $T_e \approx \SI{100}{\electronvolt}$ initially to $T_e \approx \SI{10}{\electronvolt}$ at $t = \SI{400}{\nano \second}$ after current start. In contrast, in the non-radiative case, the electron temperature remains high around $T_e \approx \SI{100}{\electronvolt}$ throughout the simulation (\autoref{fig:lineouts}d), which is much higher than in the radiatively-cooled case. We will discuss this increase in density and drop in temperature in the context of the overall pressure balance of the layer in \S \ref{sec:collapse}. Finally, we can observe plasmoids in the current sheet at $t = \SI{200}{\nano \second}$ (\autoref{fig:density}h); however, these plasmoids disappear later in time, as seen in \autoref{fig:density}i, creating a relatively homogeneous reconnection layer. 

Radiative cooling also modifies magnetic flux pile-up outside the reconnection layer. Early in time, we still observe shocks upstream of the current sheet (\autoref{fig:density}h). However, for $t > 200$ ns, pile-up is no longer mediated by a shock, as observed in {\color{blue} Figures} \ref{fig:lineouts}(e-h). Instead, there is a relatively small accumulation of magnetic flux just outside the reconnection layer (\autoref{fig:lineouts}f), while the ion density remains continuous, and only undergoes compression inside the reconnection layer. Consequently, the properties of the plasma just outside the current sheet are different compared to the non-radiative case. 

 The primary effects of radiative cooling on the structure of the reconnection layer can be summarized as follows --- (1) radiative cooling leads to a denser and thinner current sheet, indicating strong density compression; (2) the current sheet is significantly colder than in the non-radiative case; (3) the current sheet is more uniform; plasmoids that are observable initially disappear later in time; and (4) there is reduced flux pile-up outside the layer, resulting in lower magnetic field and density just outside the layer, than in the non-radiative case. We provide further discussion on these effects in the next section.

\subsection{Discussion of two-dimensional simulations}
\label{sec:discussion}

In this section, we compare and contrast the simulation results from the non-radiative and radiatively-cooled cases. In \S \ref{sec:pileup}, we discuss the decreased magnetic flux pile-up outside the layer observed in the radiatively-cooled case, which results in a lower magnetic field and density of the inflow into the current sheet. Next, we discuss the global properties of the layer in~\S \ref{sec:global_props}, and characterize differences in outflows from the reconnection layer, and in the global reconnection rate. We then discuss the radiatively-driven strong compression of the current sheet, which generates a thinner and denser layer in the radiatively-cooled simulation (\S \ref{sec:collapse}). Finally, we discuss the differences in plasmoid structure and temporal evolution between the two cases in~\S \ref{sec:plasmoids}. In the non-radiative case, plasmoids continue to grow after formation, while they collapse in the radiatively-cooled case, generating a comparatively homogenous current sheet.

\subsubsection{Magnetic flux pile-up}
 \label{sec:pileup}

In the non-radiative case, and in the radiatively-cooled case before the onset of collapse, we observe the formation of shocks on either side of the reconnection layer due to magnetic flux pile-up. Flux pile-up occurs when the rate of magnetic flux injection $\tau_\text{inj}^{-1}/\tau_A^{-1} \sim V_\text{in}/V_\text{A,1} \equiv M_\text{A,1}$ exceeds that of flux annihilation in the reconnection layer $\tau_R^{-1}/\tau_A^{-1}$ \citep{biskamp1986magnetic}. Here, $\tau_{\text{inj}}^{-1}$ and $\tau_R^{-1}$ are the flux injection and reconnection rates respectively, $\tau_A$ is the Alfvén transit time, $V_{\text{in}}$ is the inflow velocity, and $V_{A,1}$ and $M_{A,1}$ are the Alfvén velocity and Mach number in the inflow respectively. Magnetic flux accumulates outside the current sheet, resulting in a local enhancement of the inflow magnetic field and a decrease in the inflow velocity, such that the injection rate is reduced until it matches the flux annihilation rate. In incompressible sub-Alfvénic flows which satisfy the pile-up condition $M_\text{A,1} > \tau_R^{-1}/\tau_A^{-1}$, pile-up is gradual and continuous  \citep{biskamp1986magnetic}. However, in cases where the inflows are super-fast magnetosonic, flux pile-up is abrupt, and mediated by a shock upstream of the reconnection layer. The presence of shock-mediated pile-up has previously been observed in experimental studies of reconnection with high-Mach number flows \citep{fox2011fast,suttle2018ion,olson2021regulation}. 

 To estimate the jumps in density and magnetic field across the shock, we calculate the sonic $M_S = U_1/C_S, $ Alfvénic $M_A = U_1/V_A$, and fast magnetosonic  $M_{FMS} = U_1/(V_A^2 + C_S^2)^{1/2}$ Mach numbers just upstream of the shock. Here, we calculate the sonic and Alfvén speeds using $C_S = \sqrt{\gamma p / \rho}$ and $V_A = B / \sqrt{\mu_0 \rho}$ respectively, where $p$ is the thermal pressure, $\rho$ is the mass density, $\gamma = 5/3$ is the adiabatic index, $U_{1}$ is the flow velocity in the shock reference frame, and $B$ is the magnetic field strength just upstream of the shock. We use line-averaged values integrated in the $y$-direction between $|y| < L /2$ for this calculation. The outflows from the wire arrays (for the non-radiative case) are supersonic ($M_S = 4.6 \pm 0.5$), super-Alfvénic ($M_A = 1.5 \pm 0.1$), and super-fast magnetosonic ($M_{FMS}\approx 1.4 \pm 0.1$). The Mach numbers remain relatively constant in time, despite the changing density, magnetic field, and velocity of the upstream flow. The compression ratios of the line-averaged density and magnetic field across the shock, also remain relatively constant in time, as expected from the unchanging upstream Mach numbers. In the simulation, both the mass density and the magnetic field are compressed by a similar magnitude across the shock, exhibiting a compression ratio of $1.8 \pm 0.4$, consistent with ideal-MHD compression.


We can model the shock transition as a fast perpendicular MHD shock, which represents a super-fast to sub-fast transition in a system with an upstream magnetic field perpendicular to the shock normal. Solutions to the Rankine-Hugoniot jump conditions show that both the upstream magnetic field and mass density are compressed by the same ratio $r$, which can be determined from the solution of a quadratic equation \citep{goedbloed_keppens_poedts_2010}:
\begin{equation}
2(2-\gamma) r^2+\left[2 \gamma(\beta+1)+\beta \gamma(\gamma-1) M_S^2\right] r-\beta \gamma(\gamma+1) M_S^2=0
\label{eq:fast}
\end{equation}

Here, $M_S$ is the upstream sonic Mach number, and $\beta$ is the upstream plasma beta. The predicted compression ratio from \autoref{eq:fast}, using $M_S = 4.6 \pm 0.5$ and $\beta = 0.12 \pm 0.05$, is $r = 1.5 \pm 0.4$, which is consistent with the range observed in the simulation. The predicted compression ratio is slightly lower than the mean compression observed in the simulation, and may result from our assumption of a planar 1D shock which neglects the velocity component parallel to the shock caused by the radial outflows from the wire arrays. As a consequence of the flux pile-up, the downstream Alfvén Mach number, and consequently, the flux injection rate into the reconnection layer, are both reduced by a factor of $r^{-3/2}$. 

Results from the radiatively-cooled simulation show decreased flux pile-up compared to the non-radiative case after the onset of radiative collapse. This is consistent with the increase in the reconnection rate due to the strong compression of the current sheet observed in the radiatively-cooled case. We expect the flux annihilation rate to be enhanced by a factor of $A^{1/2}$ in the radiatively-cooled reconnection system \citep{uzdensky2011magnetic}. Here, $A \equiv \rho_{\text{layer}}/\rho_{\text{in}}$ is the ratio of the mass density of the reconnection layer to that just outside the layer. Because of the increased reconnection rate, a higher flux injection rate can be supported, reducing flux pile-up. A more detailed discussion of the effect of radiative cooling on the reconnection rate is provided in the next subsection. Flux pile-up modifies the plasma conditions just outside the reconnection layer, and thus must be accounted for in the analysis of experimental data before the onset of radiative collapse, or even after collapse in cases where the compression of the layer is weak enough that the flux injection rate exceeds the reconnection rate, as described later in \S \ref{sec:transport_effect}. 

\subsubsection{Lundquist Number, Outflow Velocity, and Reconnection Rate}
\label{sec:global_props}
\begin{figure} 
\includegraphics[page=5,width=1.0\textwidth]{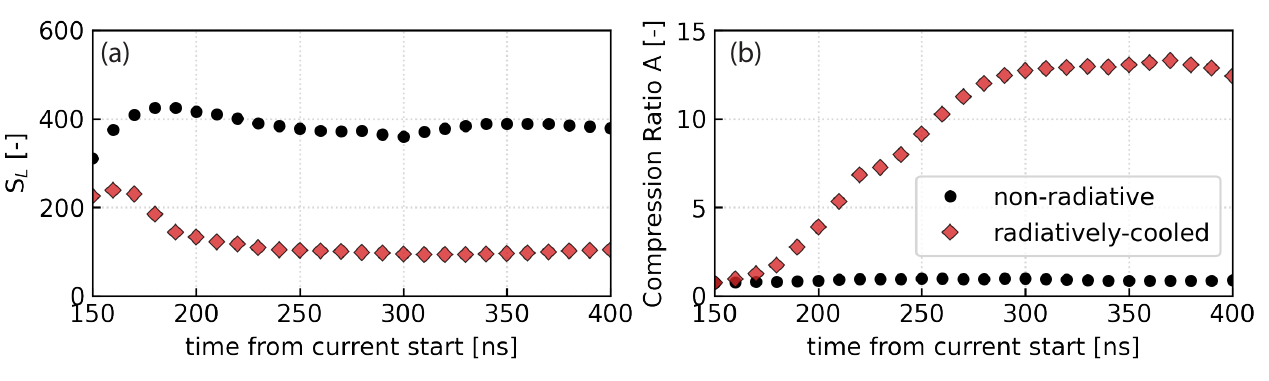}
\centering
\caption{(a) Time evolution of the Lundquist number the non-radiative (black) and the radiatively-cooled cases (red). The Lundquist number is lower for the radiatively-cooled case. (b) Time evolution of the density compression ratio of the current sheet for the non-radiative (black) and the radiatively-cooled cases (red). Results are shown for $t \geq 150$ ns, before which the layer has not fully formed.}
\label{fig:lundquist}
\end{figure}

\autoref{fig:lundquist}a compares the temporal evolution in the Lundquist number $S_L = V_{\text{A,in}}L/\bar{\eta}$ for the non-radiative and radiatively-cooled cases. Here, $V_{\text{A,in}}$ is the Alfvén speed calculated just outside the current sheet at $x = \pm 2\delta$, and $\bar{\eta}$ is the magnetic diffusivity of the layer, averaged over the  current sheet between $|x| \leq \delta$. In the non-radiative case, the Lundquist number $S_L$ increases as the current sheet forms, then reaches a relatively stationary value of $S_L \approx 400$ at $t \geq 170$ ns. For the radiatively-cooled case, the Lundquist number is similar to that in the non-radiative case early in time, but begins to fall at $t \approx 150$ ns, and reaches a steady value of $S_L \approx 100$ later in time ($t \geq 200$ ns). The change in the Lundquist number is consistent with the time of onset of radiative cooling, as observed in \S \ref{sec:radiative}. The lower Lundquist number in the radiatively-cooled case is primarily a consequence of reduced layer temperature (\autoref{fig:lineouts}h). As mentioned in \S \ref{sec:radiative}, the layer temperature falls from about \SI{100}{\electronvolt} to \SI{10}{\electronvolt} due to radiative cooling. Since the plasma (Spitzer) resistivity scales with electron temperature as $\eta \sim \bar{Z}T^{-3/2}$, a lower temperature leads to a more resistive layer, and the global Lundquist number $S_L$ becomes smaller. The average ionization $\bar{Z}$ in the current sheet also changes from approximately $11$ in the non-radiative case, to about $3.5$ in the radiatively-cooled case, but this does not compensate for the change in temperature.

In \autoref{fig:lundquist}b, we compare the density compression ratios $A \equiv \rho_{\text{layer}}/\rho_{\text{in}}$ of the current sheet for the non-radiative and radiatively-cooled cases. For the non-radiative case, the mass densities inside and outside the layer are similar, resulting in a compression ratio of $A \approx 1$. The compression ratio for the radiatively-cooled case is also about 1 early in time, but as radiative losses from the layer become more significant,  the compression ratio begins to increase around \SI{170}{\nano\second}, and approaches $A \approx 13$ later in time. The strong compression of the current sheet due to radiative cooling is indicative of radiative collapse. This occurs when an increase in compression of the layer causes radiative losses to increase faster than Ohmic dissipation \citep{uzdensky2011magnetic}. We revisit radiative collapse of the layer in \S \ref{sec:collapse}.

\begin{figure} 
\includegraphics[page=6,width=1.0\textwidth]{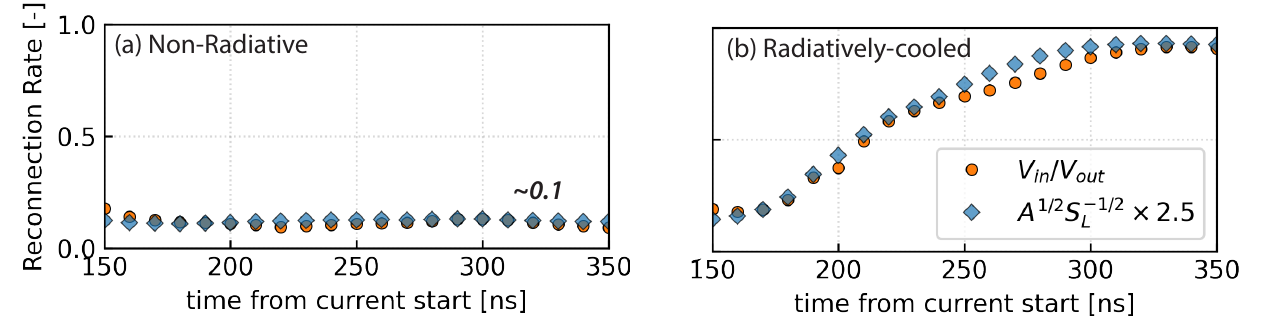}
\centering
\caption{ Reconnection rate for the non-radiative (a), and radiatively-cooled cases (b). Here, a factor of 3 is used as the constant of proportionality for the theoretical scaling. Results are shown for $t \geq 150$ ns, before which the layer has not fully formed.}
\label{fig:rate}
\end{figure}


In the non-radiative case, the ion sound speed inside the layer $C_{S,CS}$ is comparable to the Alfvén speed in the inflow $V_{\text{A,in}}$, which indicates that the magnetic tension and pressure gradient forces are roughly equal in magnitude. The outflow velocity is higher than the inflow Alfvén velocity $V_{\text{A,in}}$ and is comparable to the magnetosonic velocity (calculated from the combination of the sound and Alfvén speeds $V_{MS}^2 \equiv V_{\text{A,in}}^2 + C_{S, CS}^{2}$). Here, we calculate the outflow velocity at a distance $y = L$ from the center of the layer, averaged over $-\delta \leq x \leq \delta$ across the layer. This shows that both magnetic tension and the pressure gradient force play a role in accelerating the plasma in the reconnection layer. This effect has been observed previously in simulations \citep{forbes1991numerical}, in pulsed-power-driven experiments of carbon wire arrays \citep{hare2017formation}, and in MRX experiments where the thermal pressure upstream of the outflow region decelerates the outflows \citep{ji1999magnetic}. 

In the radiatively-cooled case after radiative collapse, the sound speed in the layer $C_{S,CS}$ is lower than the inflow Alfvén speed $V_{\text{A,in}}$ by a factor of $>2$ , consistent with the decreased layer temperature. The magnetosonic velocity is then approximately equal to the Alfvén speed $V_{MS} \approx V_{\text{A,in}}$, and the outflow velocity, therefore, agrees well with the Alfvén velocity in the inflow. The plasma is primarily accelerated by the magnetic tension of the reconnected field. Consequently, the outflow velocity is smaller in the radiatively-cooled case than in the non-radiative case, where the plasma is accelerated by both magnetic tension and pressure gradient forces. This is consistent with \cite{uzdensky2011magnetic}, which shows that unlike in usual Sweet-Parker theory, the tension force is expected to be much larger than the pressure gradient force in the radiatively-cooled case.

In \autoref{fig:rate}, we compare the reconnection rate between the two cases. We calculate the reconnection rate from the ratio of the flow velocity into the layer $V_{\text{in}}$ at $x = \pm 2 \delta$, to the outflow velocity $V_{\text{out}}$. After layer formation, the reconnection rate assumes a steady value of $V_{\text{in}}/V_{\text{out}} \approx 0.1$ for the non-radiative case. In the radiatively-cooled case, the reconnection rate increases from an initial value of about $0.1$ and reaches a value of roughly 0.9, about $9$ times higher than the non-radiative rate. For both cases, the reconnection rate is consistent with the scaling provided by compressible Sweet-Parker theory with radiative cooling, i.e. $V_{\text{in}}/V_{\text{out}} \sim A^{1/2}S_L^{-1/2}$ \citep{uzdensky2011magnetic}. In \autoref{fig:rate}, we use a constant of proportionality = $2.5$. We can attribute the high reconnection rate in the radiatively-cooled case to strong compression of the current sheet ($A \approx 13$), and to the lower Lundquist number $S_L \approx 100$ of the colder layer. 

\subsubsection{Radiative Collapse}
\label{sec:collapse}

\begin{figure} 
\includegraphics[page=7,width=1.0\textwidth]{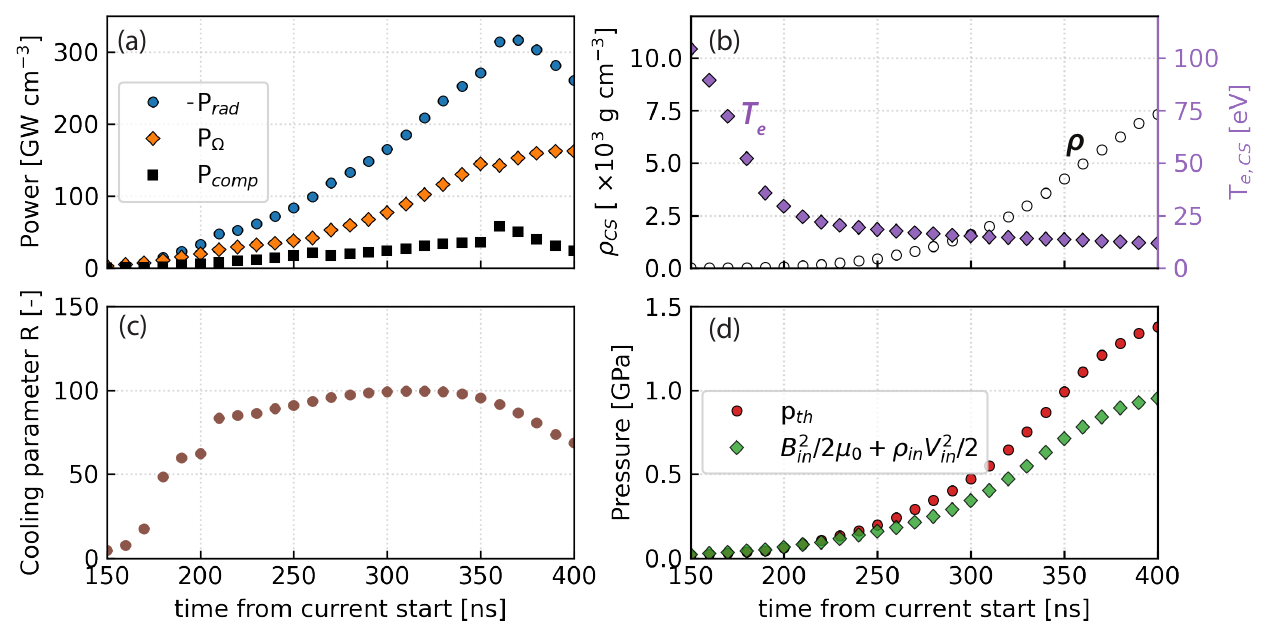}
\centering
\caption{(a) Radiative loss and Ohmic dissipation in the radiatively-cooled current sheet. (b) Temporal evolution of current sheet mass density and electron temperature. (c) Variation of the cooling parameter with time. (d) Pressure balance in the current sheet. Results are shown for $t \geq 150$ ns, before which the layer has not fully formed.}
\label{fig:radiation}
\end{figure}

The radiative collapse of the reconnection layer is characterized by a sharp decrease in its temperature, and strong compression of the layer. To understand the temporal evolution of the layer temperature, we probe the various terms in the energy equation. In our system, Ohmic and compressional heating are the dominant sources of internal energy addition to the layer, while radiative loss is the dominant loss term. Contributions of the advective terms, viscous heating, and conductive losses are comparatively small.  In \autoref{fig:radiation}a, we compare the volumetric radiative power loss $P_{rad}$, Ohmic dissipation rate $P_\Omega = \eta j^2$, and the compressional heating rate $P_{comp.} =-p(\nabla \cdot \mathbf{v})$ in the current sheet. The radiative loss from the layer is around $2$ times larger than the total heating provided by the Ohmic and compressional terms. Radiative losses are initially smaller than Ohmic dissipation right after layer formation, but begin to dominate at $t \approx 200$ ns, which is consistent with the sharp drop in the electron temperature and simultaneous density compression of the current sheet at this time, as shown in \autoref{fig:radiation}b. We quantify the relative importance of radiative loss using the cooling parameter $R_{\text{cool}} \equiv \tau_{\text{cool}}^{-1}/\tau_{A}^{-1}$, which is the ratio of the radiative cooling rate $\tau_{\text{cool}}^{-1} = (\gamma -1) P_{\text{rad}}/p_{\text{th}}$ to the Alfvénic transit rate $\tau_{A}^{-1} = V_{\text{A,in}}/L$ in the layer, as mentioned earlier in \S \ref{sec:intro}. \autoref{fig:radiation}c shows that the cooling parameter is small initially, but rises sharply between $180 < t < 200$ ns to reach a value of $R_{\text{cool}} \approx 100$. The rise in the cooling parameter is consistent with the time at which we observe radiative cooling to become significant in \S \ref{sec:radiative}. 

Finally, we compare the thermal pressure inside the current sheet $p_{\text{th}}$ with the kinetic $\rho_{\text{in}}V_{\text{in}}^2/2$ and magnetic pressures $B_{\text{in}}^2/2\mu_0$ upstream of the layer (\autoref{fig:radiation}d). The thermal pressure roughly balances the combined upstream kinetic and magnetic pressures. The thermal pressure in the layer continues to rise despite the sharp fall in the layer electron temperature. This is facilitated by the simultaneous increase in the density of the layer, as seen in \autoref{fig:radiation}b. Compression of the layer, therefore, maintains pressure balance with the upstream kinetic and magnetic pressures. 

\begin{figure}
    \centering
    \includegraphics[page=1,width=1.0\textwidth]{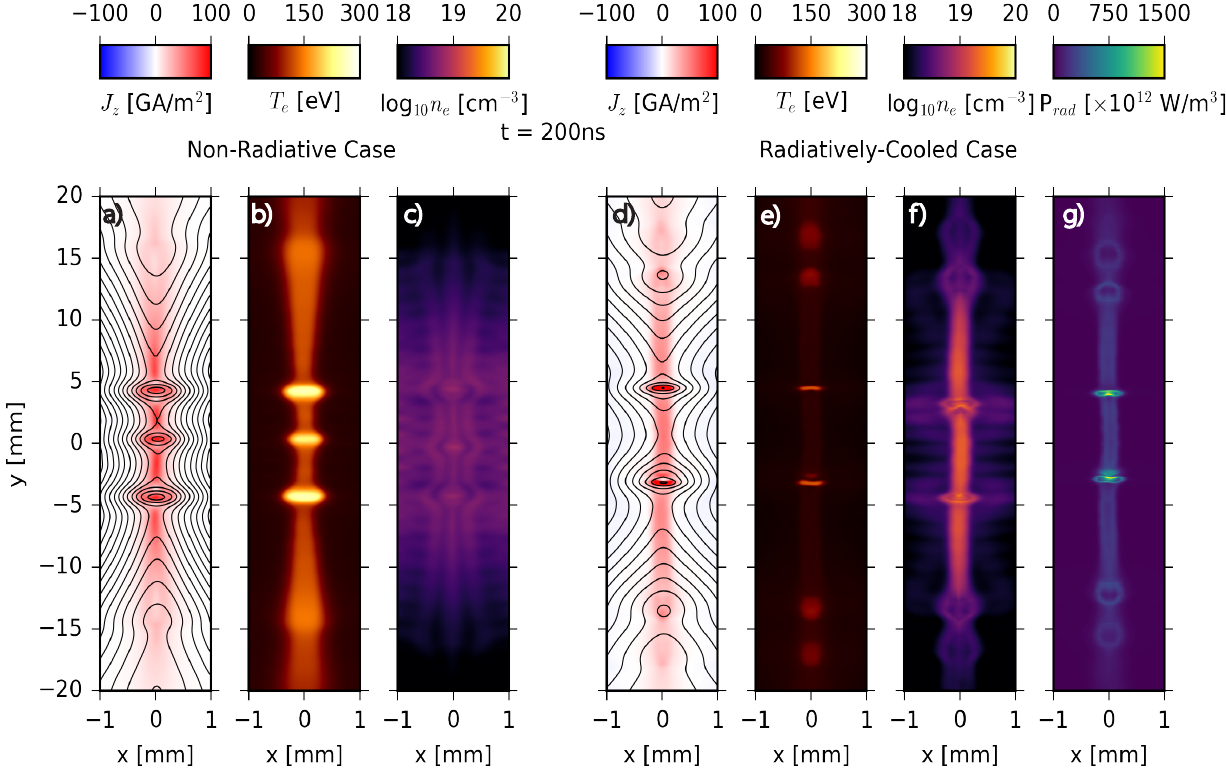}
    \caption{(a)-(c) For the non-radiative case; two-dimensional maps of (a) current density $j_z$ overlaid by contours of the magnetic vector potential $A_z$, (b) electron temperature $T_e$, (c) electron density $n_e$. (d)-(g) For the radiatively-cooled case; two-dimensional maps of (d) current density $j_z$ overlaid by contours of the magnetic vector potential $A_z$, (e) electron temperature $T_e$, (f) electron density $n_e$. (g) Radiative power loss per unit volume, as generated by XP2 (see Section \ref{sec:syntheticdiagnostics}). }
    \label{fig:2d_profile}
\end{figure}

\subsubsection{Plasmoid behaviour}
\label{sec:plasmoids}

In the previous section, we looked at the global properties of the layer by taking one-dimensional profiles, averaged along the length of the layer. However, by taking these averages, we overlook the significant modulations in the plasma properties along the $y-$direction caused by the plasmoids. Here, we look at the plasmoids in detail, and in particular we characterize the temporal evolution of the widths of the plasmoids. 

\autoref{fig:2d_profile} shows the reconnection layer at $t=200$ ns after current start, with plasmoids in both the radiatively-cooled and non-radiative cases. We choose this time as it is after the onset of radiative cooling, but before the disappearance of the plasmoids in the radiatively-cooled simulation. The findings at this time are also generally representative of later times in the non-radiative simulation.  We observe plasmoids at Lundquist numbers of 200-400, which is lower than the canonical critical Lundquist number of $S_{L,C} \sim 10^4$ \citep{loureiro2007instability}. The presence of plasmoids at these Lundquist numbers, however, is consistent with observations of plasmoids at $S_L \sim 100$ in previous pulsed-power-driven experiments \citep{hare2017anomalous}. Modulation in the inflows caused by the discrete nature of the wires may seed this instability, enabling it to occur at $S_L < S_C$. Furthermore, the system is highly compressible, strongly driven, and exhibits non-uniform resistivity, which are effects not included in the original calculation of the critical Lundquist number. Localized resistivity, in particular, is known to promote fast Petchek-type reconnection \citep{sato1979externally,biskamp2001localization,kulsrud2001magnetic,uzdensky2003petschek}, which may affect plasmoid onset and growth. 

For the non-radiative case (\autoref{fig:2d_profile}a), we plot the current density $j_z$ overlaid with the contours of the magnetic vector potential $A_z$ (which are the magnetic field lines), and we show the electron temperature $T_e$ and density $n_e$ in \autoref{fig:2d_profile}(b-c).
In this case, we see that the plasmoids (O-points~in~$A_z$) carry more current than the current sheets or layers (X-points~in~$A_z$) which separate them; the current density in the plasmoids is higher than in the sheets by a factor of about 2.  
Additionally, we observe that the plasmoids are significantly hotter than the rest of the reconnection layer (by a factor of $ >2$). The electron density in the plasmoids is also higher by a factor of $1.2-1.5$, primarily due to the pinching of  material inside the plasmoid.

We plot the same quantities for the radiatively-cooled case in {\color{blue}} \autoref{fig:2d_profile}(d-g). Here, similar to the non-radiative case, the current is localized within the plasmoids (\autoref{fig:2d_profile}d) and is higher than in the layer by a factor of about $3$. 
There is a larger separation between successive contours of the magnetic vector potential, representing a weaker magnetic field, consistent with the reduced flux pile-up observed in \autoref{fig:lineouts}. 
The plasmoids are still hotter than the rest of the layer (\autoref{fig:2d_profile}e); however, both the plasmoids and the layer are cooler than their counterparts in the non-radiative case, as expected due to radiative cooling. The layer has cooled significantly to roughly \SI{20}{\electronvolt} at this time compared to about \SI{100}{\electronvolt} in the non-radiative case, while the plasmoids have cooled from about \SI{240}{\electronvolt} in the non-radiative case to roughly \SI{75}{\electronvolt} in the radiatively-cooled case.
Lastly, the electron density in the plasmoids is about $2$ times as high as the surrounding layer, as shown in \autoref{fig:2d_profile}f. The plasmoids and current sheet are also about 3 times as dense as in the non-radiative case, as expected from the cooling-driven compression of the layer at this time.

\begin{figure}
    \centering
    \includegraphics[page=3,width=1.0\textwidth]{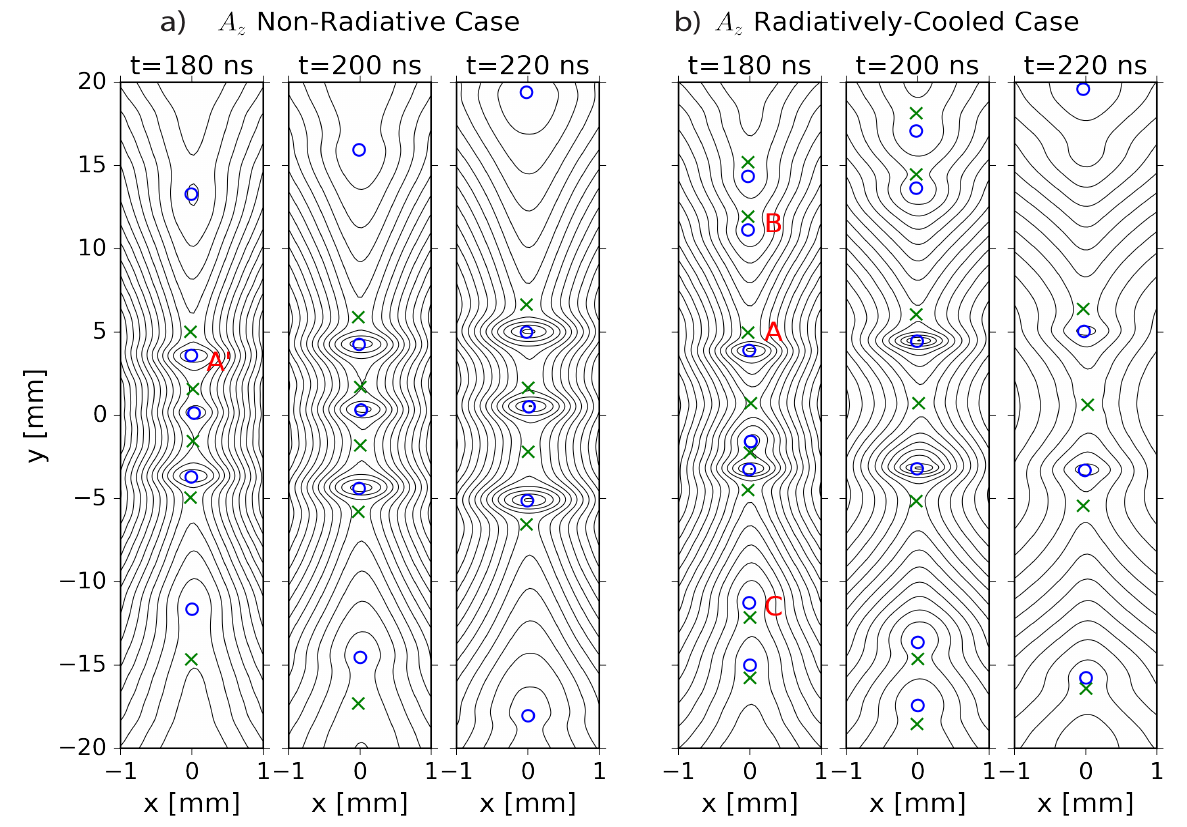}
    \caption{Contours of $A_z$ plotted at three different times ($t = 180$, $200$ and $220$ ns after current start) for (a) without radiative cooling and (b) with radiative cooling. O-points (X-points) are identified as minima/maxima (saddle points) in $A_z$ and are marked in blue (green). Specific plasmoids are labeled A' (non-radiative case) and A, B, C (radiative case), and their width is tracked in \autoref{fig:plasmoid_width}b.}
    \label{fig:x_o_points}
\end{figure}

\begin{figure}
    \centering
    \includegraphics[page=2,width=1.0\textwidth]{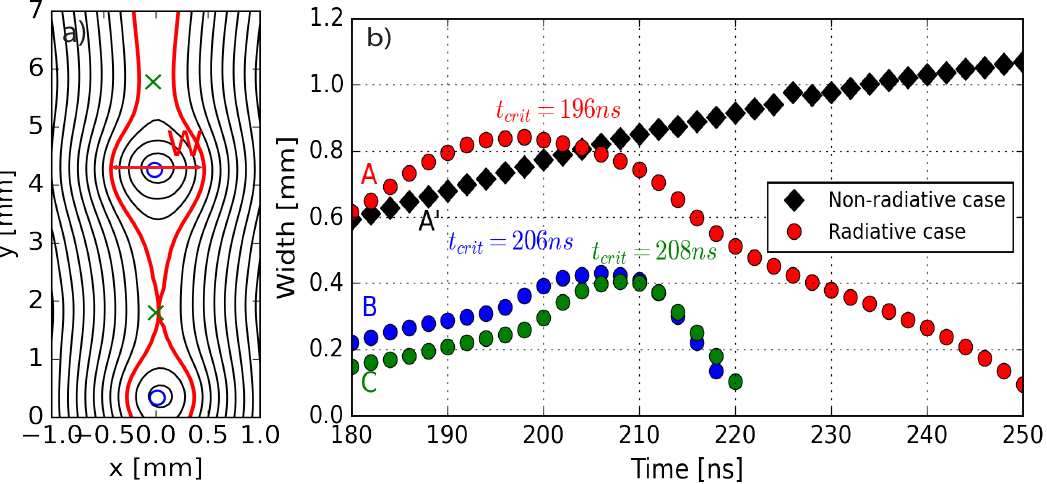}
    \caption{a) Method for calculating the width of the plasmoid; find the closest X-point, and define the width as the separation in x of the contour passing through the X-point, at the y-position of the plasmoid (O-point). b) The evolution of the width of four plasmoids; one (A') from the non-radiative case and three (A, B, C) from the radiative case. Plasmoids B and C disappear at $t \approx 210$ ns.}
    \label{fig:plasmoid_width}
\end{figure}

Since the plasmoids are both hot and dense, they are regions of strong radiative loss, as shown in \autoref{fig:2d_profile}g. The volumetric power loss rate from the plasmoids $q_{rad,p}$ is roughly an order-of-magnitude higher than that from the layer $q_{rad,L}$. By comparing the total power output from the plasmoids $\left[\sim N q_{rad,p} W^2\right]$ and the layer $\left[\sim q_{rad,L} (2L)(2\delta)\right]$, we find that power emitted from the plasmoids is roughly 1.3 times that from the rest of the layer. Here, $W$ is the plasmoid width, and $N$ refers to the number of plasmoids in the layer. To understand the role that strong radiative cooling has on the evolution of the plasmoids, we track the plasmoids and their width by finding the O-points (local maxima) of $A_z$. We also identify the X-points, or the magnetic null points, by finding the saddle points of $A_z$. We mark the X-points and the O-points on contours of $A_z$ in \autoref{fig:x_o_points}, for both the non-radiative and radiatively-cooled cases at several successive time snapshots. We see that for both cases, the plasmoids move along the $y$-axis with the outflows from the reconnection layer. We note that for the radiatively-cooled case shown in figure \autoref{fig:x_o_points}b, two plasmoids at around $y=-3$ mm coalesce between 180 ns and 200 ns. We define the plasmoid width as the horizontal separation at the O-point between the magnetic separatrix contour passing through the nearest X-point, shown graphically in red in \autoref{fig:plasmoid_width}a.
In \autoref{fig:plasmoid_width}b, we compare the change of the width in time of four plasmoids: one (A') from the non-radiative case, and three (A, B, and C) from the radiatively-cooled case. The plasmoid labeled A' corresponds to the largest plasmoid present in the non-radiative case (black diamonds), A is the corresponding plasmoid in the radiatively-cooled case (red circles). B and C are smaller plasmoids in the radiatively-cooled case (blue and green circles).

For plasmoid A', we see that the plasmoid width increases monotonically with time, which is consistent with the injection of magnetic flux and mass density into the plasmoid from the neighboring X-points. For the radiatively-cooled case, however, plasmoid A initially grows faster than A' and reaches a larger width, but then begins to shrink.
A similar trend is observed for plasmoids B and C; initially, there is an increase in plasmoid width, followed by a decrease.
We define the time at which an individual plasmoid reaches its maximum size and begins to shrink as $t_{crit}$, such that $d W/d t|_{t_{crit}} = 0$. For plasmoid A, $t_{crit}=196$ ns, and this plasmoid eventually disappears at $t=250$ ns. For plasmoids B and C, $t_{crit} \approx 207$ ns, and thus we observe that the smaller plasmoids collapse at a later time. All of these critical times occur around the time $t\approx200$ ns at which globally we observe that the volumetric radiative cooling rate $P_{rad}$ becomes comparable to the Ohmic heating rate $P_{\Omega}$ (\autoref{fig:radiation}a). 

In summary, we find that plasmoids begin to collapse when radiative cooling becomes important, and are eventually extinguished. 
This indicates an interplay between the tearing instability and the cooling instability \citep{VanHoven1984, schoeffler2023high}. The mechanism behind this coupling will be investigated in greater detail in a future publication, using simulations with simpler geometries and boundary conditions, rather than the entire experimental domain. 

\subsection{Effect of Radiation Transport}
\label{sec:transport_effect}

\begin{table*}\centering
\ra{1.3}
\caption{Comparison of results between the non-radiative and radiatively-cooled cases at peak current ($t = 300$ ns) for the two-dimensional simulations. At $t = 300$ ns, the reconnection layer in the radiatively-cooled cases (local loss and radiation transport models) has collapsed.}
\begin{tabular}{cccc}
\toprule
\multirow{2}{*}{Quantity} & \multirow{2}{*}{Non-Radiative} & Radiatively-Cooled & Radiatively-Cooled \\
& & (Local Loss) & (Radiation Transport) \\
\midrule
Reconnection Layer Length $2L$ (\SI{}{\milli \meter}) & 35 & 37 & 39 \\
Reconnection Layer Width $2\delta$ (\SI{}{\milli \meter}) & 0.6 & 0.2 & 0.3 \\
Aspect ratio $L/\delta$  & 63 & 150 & 120 \\
Reconnection Layer Temperature $T_e$ (\SI{}{\electronvolt}) & 80 & 15 & 18 \\
Lundquist Number $S_L = V_{A,in}L/\bar{\eta}$  & 360 & 95 & 80 \\
Compression Ratio $A = \rho_{L}/\rho_{\text{in}}$  & 1 & 13 & 6 \\
Inflow Velocity $V_{\text{in}}$ (\SI{}{\kilo \meter \per \second}) & 16 & 58 & 50\\
Outflow Velocity $V_{\text{out}}$ (\SI{}{\kilo \meter \per \second}) & 121 & 67 & 56 \\
Inflow Magnetic Field $B_{\text{in}}$ (T) & 38 & 18 & 13 \\
Inflow Alfvén Velocity $V_{\text{A,in}}$ (\SI{}{\kilo \meter \per \second}) & 81 & 46 & 37 \\
Normalized Reconnection Rate (-) & 0.1 & 0.9 & 0.9 \\
\bottomrule
\end{tabular}
\label{tab:table}
\end{table*}

To investigate the effects of radiation transport on the reconnection process, we repeat the 2D simulation with the P$_{1/3}$ multi-group radiation transport model \citep{crilly2022spk}. This 2D simulation was run with the same array parameters, resolution, and initial and boundary conditions as the radiatively-cooled simulation with the local loss model described in \S\ref{sec:radiative}. The initial wire core temperature, however, was increased to 0.25~eV (from 0.125~eV in the local loss model results shown above). This increased core temperature does not make a significant difference to the local loss model simulations, and was chosen to better reproduce existing experimental results.

The global reconnection dynamics observed with radiation transport are similar to those with local loss. A reconnection layer forms at the mid-plane ($x =\SI{0}{\milli\meter}$) between the wire arrays, and magnetic flux pile-up generates shocks on either side of the layer. \autoref{tab:table} compares key properties in the reconnection layer and the inflow between the local loss and radiation transport simulations at \SI{300}{\nano\second}, by which time the layer has collapsed in both simulations. The layer temperature in the radiation transport simulation ramps up to about 120~eV, before beginning to drop around 160~ns due to radiative collapse. By 300~ns, the layer temperature falls to 18~eV, which is slightly higher than in the local loss simulation ($T_{layer} \approx \SI{15}{\electronvolt}$). Cooling is accompanied by strong compression of the reconnection layer in the radiation transport case, similar to the local loss case. The compression ratio rises from $A \approx 1$ before radiative collapse to about 6 after collapse (around 300~ns). The compression is about 2.2 times lower than that in the local loss simulation ($A \approx 13$) at this time (see \autoref{fig:lundquist}). Magnetic flux pile-up is also observed to persist longer in the radiation transport simulation. Shocks disappear in the local loss simulation by 250~ns (see \autoref{fig:lineouts}), while the shocks begin to disappear around 300~ns in the radiation transport simulation. The presence of magnetic flux pile-up and shocks is consistent with the lower compression and reconnection rate in the radiation transport simulation at 250~ns [$V_{in}/V_{out} (t = \SI{250}{\nano\second}) \approx 0.3, \, A \approx 3$], than in the local loss case [$V_{in}/V_{out} (t = \SI{250}{\nano\second}) \approx 0.7, \, A \approx 9$]. Later at around 300~ns, the reconnection rate becomes similar in both cases, as shown in \autoref{tab:table}.

To understand the lower compression in the radiation transport simulation, we explore the pressure balance between the layer and inflow. Similar to the local loss simulation (see \autoref{fig:radiation}d), the layer pressure still balances the combined magnetic and kinetic pressures outside the layer. However, the total pressure outside the layer is roughly $2$ times lower in the radiation transport simulation than in the local loss case. At 300~ns, the total inflow pressure in the local loss simulation is about 500~MPa, while in the radiation transport simulation, it is about 250~MPa. The lower inflow pressure in the radiation transport simulation explains the weaker compression of the reconnection layer after radiative collapse. 

\begin{figure}
    \centering
    \includegraphics[page=3,width=1.0\textwidth]{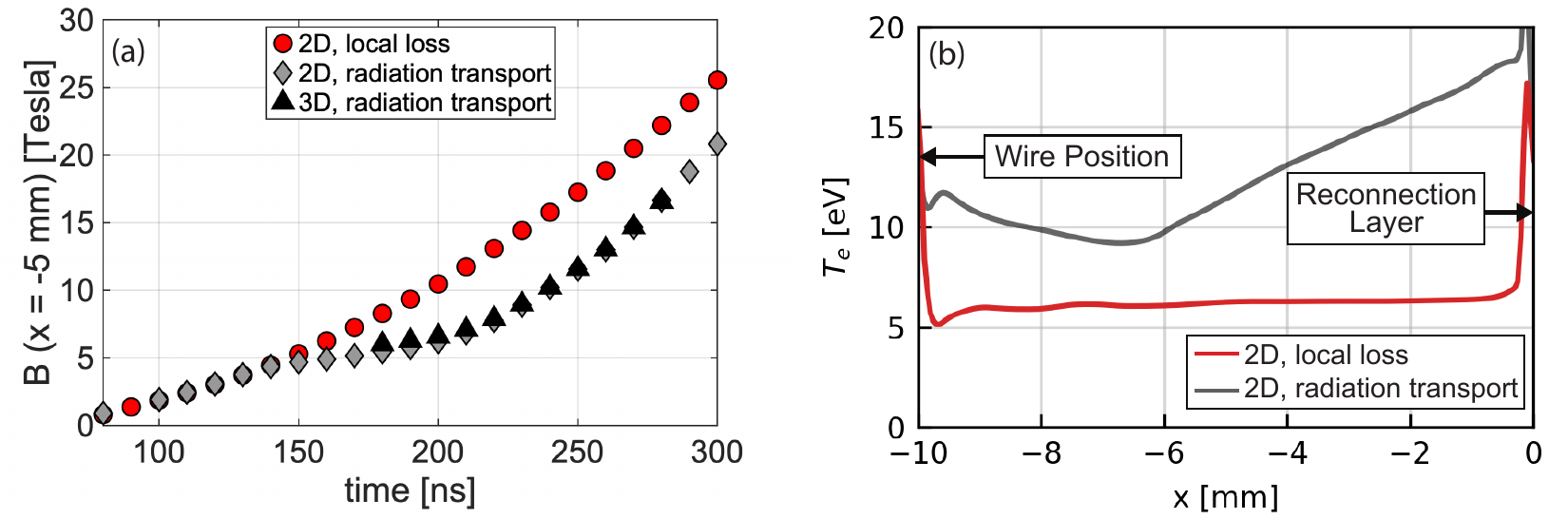}
    \caption{(a) Comparison of the magnetic field strength at 5 mm from the wires for the 2D simulation with local radiation loss (red circles), 2D simulation with multi-group radiation transport (grey diamonds), and 3D simulation with multi-group radiation transport (black triangles). Results are shown for $t > \SI{100}{\nano \second}$ to account for the transit time. For the 3D simulations, we only show output between 180 - 280~ns. (b) Variation of electron temperature along the $x-$axis between the wire position ($x = \SI{-10}{\milli\meter}$) and the reconnection layer ($x = \SI{0}{\milli\meter}$) at 300~ns for the 2D local loss (red) and 2D radiation transport (grey) simulations.} 
    \label{fig:rad_transport}
\end{figure}

The reduced inflow pressure is a consequence of lower advected magnetic field and flow velocity in the plasma ablating from the wire arrays in the radiation transport simulation. The lower magnetic field and velocity not only generate a lower pressure in the post pile-up region, but also result in decreased Alfvén $B_{in}/\sqrt{\mu_0\rho_{in}}$ and inflow velocities $V_{in}$, as shown in \autoref{tab:table}. Consistent with the lower $V_{A,in}$, the outflow velocity $V_{out}$ is also lower in the radiation transport simulation. \autoref{fig:rad_transport}a compares the advected magnetic field at a distance of 5~mm from the wires for the radiation transport (grey) and local loss simulations (red). The magnetic field is initially similar in both cases but begins to deviate around 150~ns. Between 150-200~ns, the magnetic field is almost constant (around 5~T), despite the increase in the driving magnetic field inside the arrays. The magnetic field begins to rise again after 200~ns at a rate similar to that in the local loss case. However, the magnitude remains lower in the radiation transport simulation than in the local loss case. After 150~ns, the velocity in the radiation transport simulation is also lower than in the local loss case.

The reduced advected magnetic field occurs due to a modification of the wire ablation dynamics, caused by heating of the wire cores in the radiation transport simulation. In the local loss case, the wire cores cool slightly with time (from about 0.25~eV to 0.2~eV between 100-200~ns). In the radiation transport simulation, however, the wire core temperature, which is initially 0.25~eV, rises significantly after 50~ns due to the re-absorption of emission from the plasma around each core, and becomes about 0.6-1.2~eV between 100-150~ns, much higher than in the local loss case. The transport of the magnetic field from inside the array to outside the array depends on the resistive diffusion rate $\tau_{\text{diff.}}^{-1} \sim \bar{\eta}_{\text{core}}/d_{\text{core}}^2$ of the field through the wire cores. Here, $d_{\text{core}} \approx \SI{0.4}{\milli\meter}$ and $\bar{\eta}_{\text{core}}$ are the wire core diameter and magnetic diffusivity respectively. The higher core temperature decreases the resistive diffusion rate by a factor of $>10$, contributing to the decreased magnetic field outside the array, which is then advected away from the wires by the plasma flow.

Finally, in addition to modifying the ablation of plasma from the wires, radiation transport also results in heating of the plasma upstream of the reconnection layer. \autoref{fig:rad_transport}b shows the variation of electron temperature along the $x-$axis between the wire position ($x = \SI{-10}{\milli\meter}$) and the reconnection layer ($x = \SI{0}{\milli\meter}$) at 300~ns, for the local loss (red) and radiation transport (grey) simulations. The temperature is lower (about 6~eV) in the local loss simulation, and spatially uniform between the wire and reconnection layer positions compared to the radiation transport simulation. In contrast, re-absorption of emission from the reconnection layer heats the plasma adjacent to the layer, and emission from the wires heats the plasma close to the wires, causing increased temperatures (10-18~eV) in the radiation transport simulation. These effects further underscore the importance of radiation transport and optical depth in these experiments.

\section{Three-dimensional simulations}
\label{sec:three-dimensionaleffects}

The two-dimensional simulations described in the previous section provide a detailed picture of the effects of radiative cooling and radiation transport in our experiment. 
To study three-dimensional effects, and to more closely predict the dynamics of the actual experiment, we extend our simulation by $\qty{36}{\milli \meter}$ (720 grid cells) in the $z$ direction. We randomly perturb the initial temperature of the wire cores along the $z$ direction to seed the axial non-uniformity in wire array ablation observed in experiments \citep{chittenden2004x}.
The 3D simulation uses a $P_{1/3}$ multi-group radiation transport model, similar to the 2D simulation described in \S \ref{sec:transport_effect}. 
Apart from these changes, all other parameters remain consistent with those used in the 2D simulations. 
Due to the high computational cost and the large size of the simulation output, we run the three-dimensional simulations until 280~ns after current start, by which time the reconnection layer has collapsed. Furthermore, we only simulate the radiatively-cooled case in 3D. In the following subsections, we provide a preliminary look at the results from the 3D simulations. A detailed description of 3D results will be provided in a future publication.

\begin{figure}
    \centering
    \includegraphics[page=1,width=1.0\textwidth]{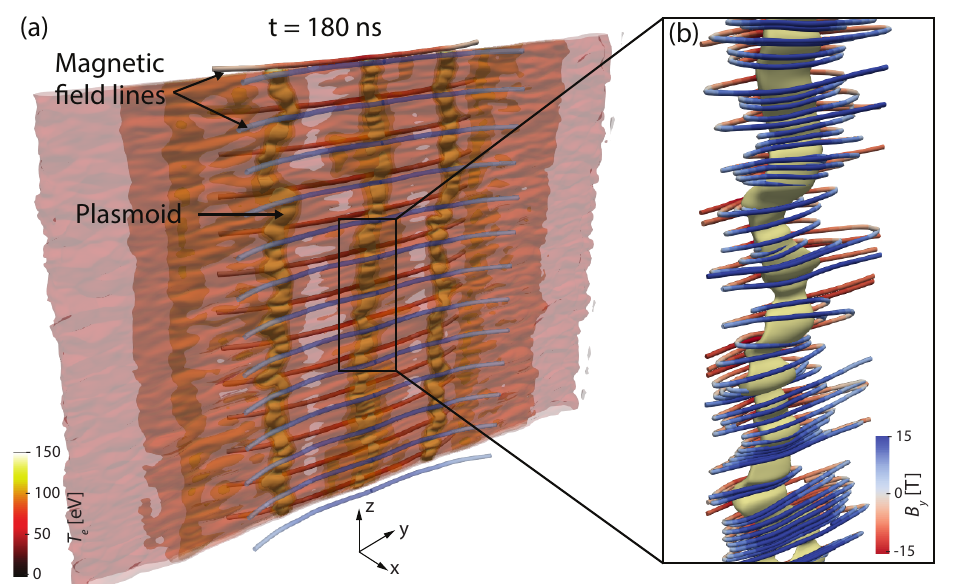}
    \caption{(a) Three-dimensional simulated electron temperature map together with the reconnecting magnetic field lines at 180 ns after current start. Plasmoids appear as columns of enhanced temperature within the reconnection layer. (b) Magnified image of a plasmoid and its magnetic field lines. The plasmoids exhibit strong kinking along the axial direction.  }
    \label{fig:3D}
\end{figure}

\subsection{Results}

To examine the global dynamics of the reconnection process, we perform the same analysis used in \S \ref{sec:results}-\S \ref{sec:transport_effect} at multiple $z$-slices in the simulation domain. The reconnection dynamics observed in 3D closely resemble the 2D radiation transport simulation (\S \ref{sec:transport_effect}), and do not vary significantly along the axial direction. \autoref{fig:rad_transport}a shows the advected magnetic field at a radial location of \SI{5}{\milli\meter} from the wires ($x = \SI{5}{\milli\meter},\, y, z = \SI{0}{\milli\meter}$) in the 3D simulation. The magnetic field closely agrees with that in the 2D radiation transport simulation. Other quantities, such as the flow velocity, temperature, and ion density in the 3D simulation are also similar to those in the 2D simulation (\S \ref{sec:transport_effect}). The temperature of the reconnection layer drops from roughly \SI{100}{\electronvolt} initially to about \SI{18}{\electronvolt} later in time, accompanied by density compression of the layer and an accelerated reconnection rate, similar to what was seen in 2D. During the compression process, the layer maintains pressure balance with the upstream magnetic and kinetic pressures. The compression ratio is initially similar in the 2D and 3D simulations; however at 280~ns, the compression ratio in the 3D simulation becomes slightly lower ($A_{3D} \approx 3$) than that in the 2D case ($A_{2D} \approx 4$). Consequently, the reconnection rate in 3D at this time is also slightly lower by a factor of about 1.1, consistent with the $V_{\text{in}}/V_{\text{out}} \propto A^{1/2}$ scaling \citep{uzdensky2011magnetic}.

\begin{figure}
    \centering
    \includegraphics[page=2,width=1.0\textwidth]{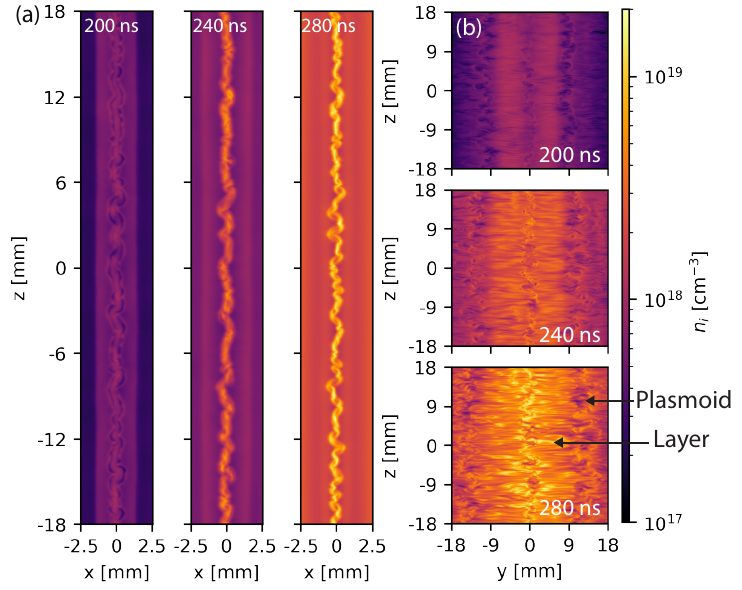}
    \caption{(a) Cross-sectional view of the electron density in the central plasmoid at the $y = 0$ $xz$-plane at 200, 240, and 280 ns. (b) Cross-sectional view of the electron density in the reconnection layer at the $x = 0$ $yz$-plane at 200, 240 and 280 ns.  }
    \label{fig:3D_ne}
\end{figure}

\autoref{fig:3D}a shows 3D electron temperature contours of the reconnection layer at \SI{180}{\nano\second} after current start, together with the reconnecting magnetic field lines upstream of the layer. Consistent with the 2D simulations, the layer is unstable to the plasmoid instability, and flux ropes (3D analogues of 2D plasmoids) appear as columns of higher-temperature plasma (about \SI{140}{\electronvolt}, yellow) compared to the rest of the layer, which exhibits a mean temperature of roughly \SI{90}{\electronvolt} (orange and red) at this time. \autoref{fig:3D}b shows a magnified view of the central flux rope and its local magnetic field topology. As expected, the field lines wrap around the plasmoid to form a magnetic flux rope. The flux ropes also exhibit helical perturbations that resemble the $m=1$ MHD kink mode. 

\subsection{Discussion of three-dimensional simulations}
\label{sec:three_dimen_discussion}

The results in 3D are consistent with those in 2D, which is expected due to the quasi-2D nature of the experiment. We observe a slight decrease in the compression ratio in the 3D simulation compared to the 2D case around 280~ns. A reduction in compression between 2D and 3D in radiative reconnection was also observed and investigated by \cite{schoeffler2023high}, for the case of relativistic reconnection with synchrotron cooling. In those simulations, the weaker compression occurred because as the magnetic field compressed the plasma, the plasma was free to move to regions of lower magnetic field along the out-of-plane $z-$direction. This was facilitated by modulations along the $z-$direction generated by the kink instability \citep{schoeffler2023high}.

The kink instability of the flux ropes appears in our simulation as early as 150 ns, when the reconnection layer has just formed. In the absence of a guide field, flux ropes exhibit the magnetic field topology of a z-pinch (see \autoref{fig:3D}b), and therefore have unfavorable MHD stability \citep{biskamp1996magnetic,biskamp1991algebraic,freidberg_2014}. The MHD kink instability of flux ropes has also been observed in other three-dimensional simulations of magnetic reconnection \citep{lapenta2011spontaneous,schoeffler2023high}. Strong radiative emission from the plasmoids can also make them susceptible to thermal cooling instabilities \citep{field1965thermal,somov1976physical}.

\autoref{fig:3D_ne}a shows the ion density at the $y=\SI{0}{\milli\meter}$ cross-section of the central flux rope in $xz$-plane at three different times (200, 240 and 280~ns). This flux rope remains at the center of the layer ($y = 0$) during the simulation, while the other two flux ropes visible in \autoref{fig:3D}a are advected away from the center of the layer by the outflows. As observed in \autoref{fig:3D_ne}a, the amplitude and wavelength of the instability remain invariant in time, which indicates saturation of the kink mode. The dominant amplitude of the kink mode in the $xz$-plane is roughly \qty{400}{\micro \meter}, and the wavelength is about \qty{2}{\milli \meter}. In \autoref{fig:3D_ne}b, we plot the cross-section of the current sheet in the $x = 0$ $yz$-plane at the same three times. In this plane, the amplitude of the modulations appears to grow with time, but this is primarily due to the velocity gradient in the outflows from the reconnection layer. The flow velocity $V_y$ increases with distance $|y|$ from the center of the layer, and becomes comparable to the Alfvén speed at $y = \pm L$, consistent with acceleration driven by the magnetic tension of the reconnected field lines. \autoref{fig:3D_ne}b also shows elongated (along $y$) modulations of the electron density in the reconnection layer in the $z-$direction. These modulations appear due to non-uniformity (along $z$) in the wire array ablation, which is seeded by modulating the initial temperature of the wire cores. The 2~mm wavelength of the flux-rope kink mode is much larger than that of the axial non-uniformity in the ablation flows ($\approx 100-\SI{300}{\micro \meter}$).

Between 150-180 ns, the Alfvén crossing time $\tau_{A,pl} = W/V_A$ (the ratio of the plasmoid width~$W$ to the Alfvén velocity, and the time scale on which MHD instabilities grow), is roughly $2-\qty{4}{\nano \second}$, while the radiative cooling time (the time scale on which cooling instabilities grow) is $\tau_{\text{cool}} \approx \qty{10}{\nano \second}$. 
Not only is the cooling time longer than the Alfvén time right after layer formation, but Ohmic and compressional heating are also stronger than radiative cooling at this time, consistent with \autoref{fig:radiation}a. 
Thus, we expect MHD instabilities, as opposed to cooling instabilities, to dominate and drive the dynamics of the layer right after its formation. 
Later in time, during the onset of radiative collapse, the cooling time $\tau_{\text{cool}} \approx \qty{1}{\nano \second}$ becomes comparable to the Alfvén crossing time. For a homogenous optically thin 1D system, the stability criteria and the growth rates of thermal cooling instabilities derived from linear theory typically depend on the derivatives of the cooling function with respect to density and temperature \citep{field1965thermal}. However, for our inhomogeneous highly-dynamic configuration with optically thick radiative emission, analytical results do not exist.
The interplay of thermal cooling instabilities and MHD instabilities of the current sheet will be a topic for further investigation.

\section{Synthetic X-Ray Diagnostics}
\label{sec:syntheticdiagnostics}

We post-process the MHD simulations using the X-ray Post Processor (XP2) code \citep{crilly2022spk} to produce synthetic X-ray images, time histories, and spectra relating to the experimental diagnostics of the MARZ experiments. 
These diagnostics are key for measuring both the radiative collapse of the reconnection layer and the formation of plasmoids. 
Time-resolved X-ray images from above the layer are spectrally-filtered to highlight dynamics of the hot, strongly emitting plasmoids. 
The aluminum K-shell line spectra contain lines from different ionization stages of Al which provide information on the temperature of the emitting plasma. 
We expect optical depth effects to be significant in the MARZ experiments given the large length of the reconnection layer, and to modify the relative intensity of Al He-$\alpha$ resonance to inter-combination lines.

\begin{figure} 
\includegraphics[page=1,width=1.0\textwidth]{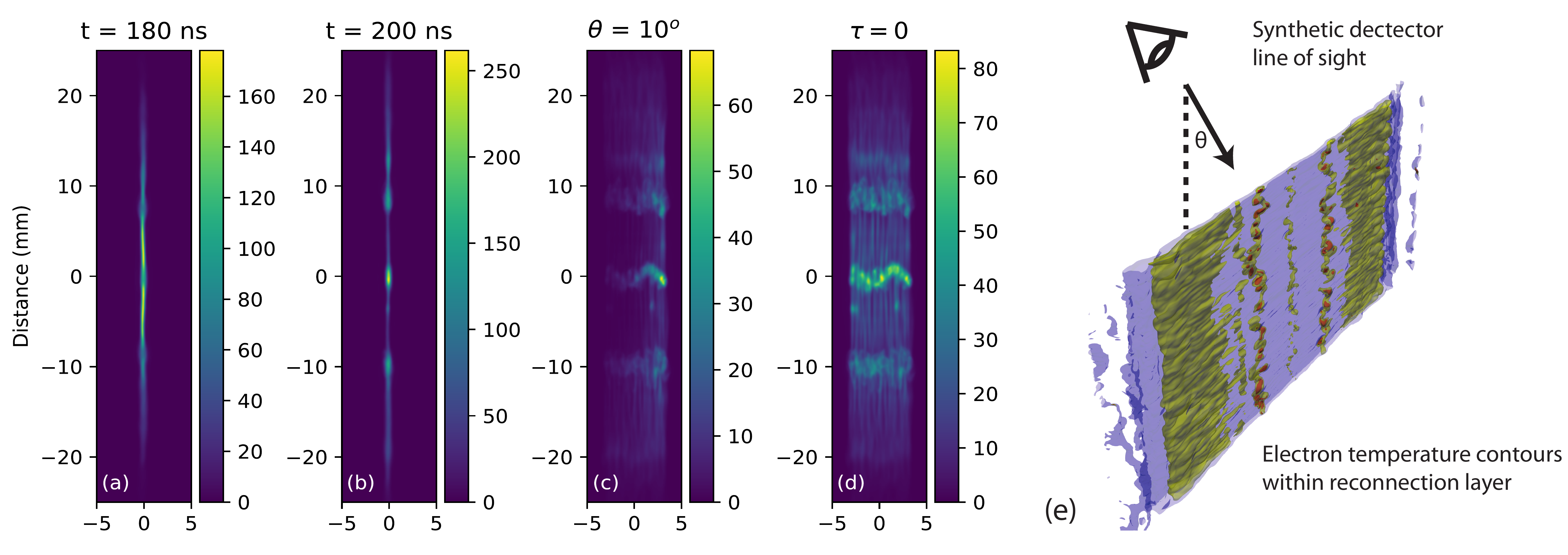}
\centering
\caption{Subplots (a) and (b) are synthetic time-resolved X-ray images filtered by 2$\mu$m of Mylar at 180 and 200 ns respectively in TW/m$^2$/sr. These were produced with the XP2 code and SpK NLTE emissivity and opacity tables. The synthetic detector line of sight is parallel with the $z$-axis. Subplots (c) and (d) are synthetic time-resolved X-ray images at 200 ns with a 10 degree offset on the detector line of sight, with and without optical depth effect respectively. The offset is such that the right hand side of the image is due to emission from the part of the layer nearest the detector. (e) A graphic showing the orientation between the layer and the synthetic detector. The layer is shown through electron temperature contours (at 30, 50, and \SI{80}{\electronvolt}) at 210 ns.}
\label{fig:MARZ_imaging_comparison}
\end{figure}

Spatially resolved X-ray spectral intensity at the detector plane is produced using multi-group SpK tables and XP2 from the 2D and 3D MHD simulations. Henke X-ray transmission data \citep{henke1993x} is used to spectrally filter the incident intensity to produce synthetic images. Images can be created along experimental diagnostic lines of sight which may not line up with the simulation grid axes, as is often the case in a realistic experimental geometry. \autoref{fig:MARZ_imaging_comparison} presents post-processed filtered X-ray images from 3D MARZ simulations, showing time evolution, viewing angle, and optical depth capabilities.

\begin{figure} 
\includegraphics[page=1,width=1.0\textwidth]{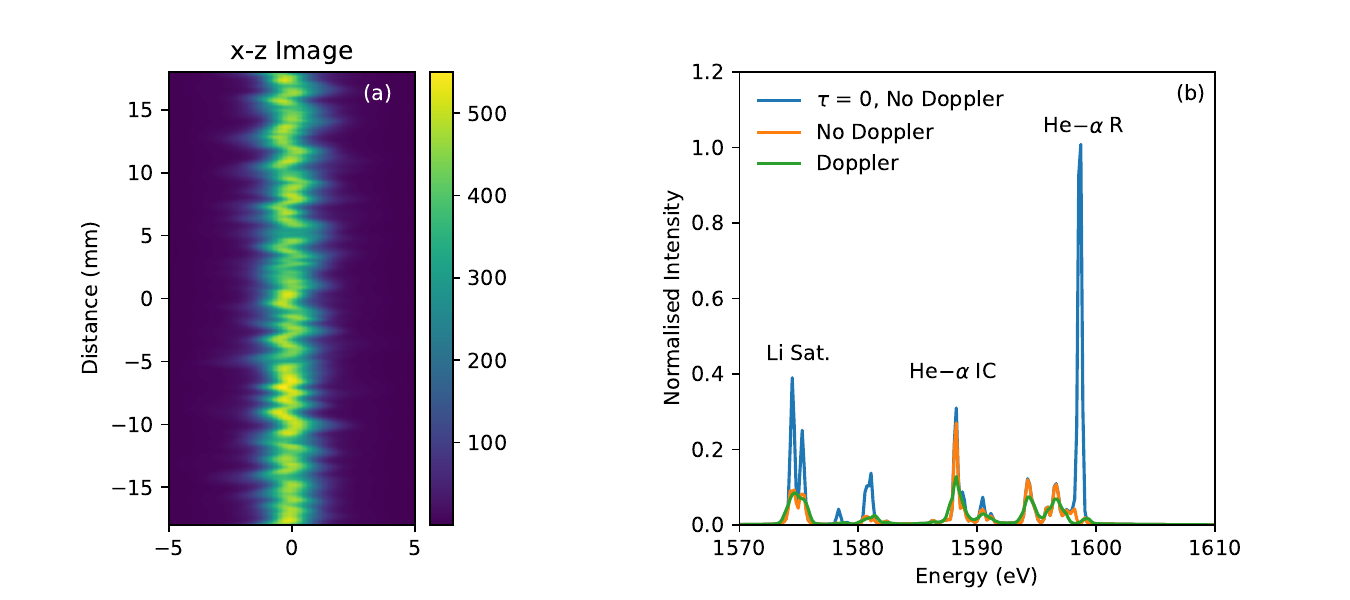}
\centering
\caption{(a) A time-integrated 2um Mylar filtered X-ray image along the layer showing perturbations in the layer structure in the inflow direction. (b) Time-integrated X-ray spectra from XP2 using SCRAM emissivity and opacity tables. Key spectral lines are labeled where R denotes resonance, IC intercombination, and Sat. satellites. }
\label{fig:MARZ_spectra_comparison}
\end{figure}

To produce accurate synthetic X-ray line spectra, spectroscopic-quality emissivity and opacity data is required, and therefore results from the SCRAM code \citep{hansen2014detailed,hansen2007hybrid} are used in XP2. The SCRAM calculations were performed at various densities and temperatures expected within the layer in simulation. The steady-state rate equations were solved including the effect of photo-pumping; SCRAM does this by assuming a cylindrical homogenous plasma with a characteristic length scale (diameter) of \SI{1}{\milli\meter}. A length scale of \SI{1}{\milli\meter} was chosen because it is comparable to the reconnection layer width, and thus the mean chord of the radiation escaping the reconnection layer. From this, SCRAM produced emissivity and opacities within the energy range of interest (1560 - \SI{1610}{\electronvolt}) with detailed term accounting and spectral line broadening effects. XP2 used the SCRAM tables to perform radiative transfer across the simulation domain and the spectra are temporally integrated over multiple output MHD time steps to model the time-integrated nature of the X-ray spectrometer available for these experiments. 
Optical depth is especially important for the amplitude of the spectral lines, and in particular, for the He-$\alpha$ resonance line, as shown in \autoref{fig:MARZ_spectra_comparison}. The presence of other lines gives information on the temperature of the emitting region --- for example, Li-like satellites are present rather than H-like lines due to the cooler plasma. 
Doppler shifts produced by the ejection of material out of the reconnection layer produce additional broadening on the spectral lines beyond that provided by Stark and thermal Doppler broadening.

These simulated measurements demonstrate XP2 and GORGON's capability to produce synthetic diagnostics which can be utilized in the experimental design and data analysis stages. They also highlight the impact of various physical processes on diagnostic signals, such as optical depth. This will aid in the interpretation of experimental results and in the identification of key signatures of magnetic reconnection and radiative collapse.

\section{Conclusions}

We performed two- and three-dimensional resistive-MHD simulations of radiatively-cooled magnetic reconnection in a pulsed-power-driven dual wire array load. These simulations elucidate the physics of the MARZ experiments, which are designed to study the effects of radiative cooling on magnetic reconnection driven by the Z pulsed-power machine. In our simulations, the arrays generate magnetized supersonic ($M_S = 4-5$), super-Alfvénic ($M_A \approx 1.5$), and super-fast magnetosonic ($M_{FMS} \approx 1.4$) flows which interact in the mid-plane to generate a radiatively-cooled current sheet.

In two dimensions ($xy$), we performed simulations without radiative cooling (non-radiative case) and with radiative cooling implemented using a local loss model (radiatively-cooled case, see \S \ref{section:radmodels}). The results at 300~ns after current start (at peak current) are summarized in \autoref{tab:table}. As described in \S \ref{sec:results}, radiative cooling results in a significantly colder layer compared to the non-radiative case. Because of the lower temperature, the Lundquist number of the reconnection layer is also smaller. Furthermore, the layer is thinner, and exhibits strong compression in the radiatively-cooled case, with a maximum density compression ratio of about $13$, as described in \S \ref{sec:global_props}. The sharp decrease in the layer temperature, together with the strong compression of the layer, is consistent with radiative collapse of the current sheet.



A comparison of the Ohmic dissipation rate and the radiative power loss shows that radiative losses exceed the rate at which magnetic energy is dissipated, causing the layer to lose internal energy faster than it can be added by Ohmic heating (see \S \ref{sec:collapse}). The strong compression results from a pressure balance across the reconnection layer --- the thermal pressure in the current sheet balances the combined magnetic and kinetic pressures outside the layer --- and consequently the density increases as the temperature drops, further increasing the rate of radiative cooling. As a consequence of the strong compression and lower Lundquist number, the global reconnection rate $V_{\text{in}}/V_{\text{out}} \approx 0.9$  is 9 times higher than in the non-radiative case, consistent with the theoretical scaling $\sim A^{1/2}S_L^{-1/2}$ predicted from compressible Sweet-Parker theory \citep{uzdensky2011magnetic}. This faster reconnection dissipates the piled-up magnetic flux, removing the magnetically-mediated shocks upstream of the reconnection layer, which are observed in the non-radiative case (see \S \ref{sec:pileup}).

In both the radiatively-cooled and non-radiative cases, the current sheet is unstable to the plasmoid instability. The plasmoids exhibit a higher density and temperature than the rest of the layer, and therefore appear as hotspots of enhanced radiative emission within the layer, as shown in \S \ref{sec:plasmoids}. In the radiatively-cooled case, the plasmoids are quenched before ejection from the layer --- the width of the plasmoids begins to decrease with time when the radiative cooling rate becomes comparable to the Ohmic dissipation rate. 

We further explore the effects of finite optical depth by implementing $P_{1/3}$ multi-group radiation transport in the 2D simulation (\S \ref{sec:transport_effect}). The results from this simulation are tabulated in the third column of \autoref{tab:table}. Radiation transport significantly modifies the ablation dynamics of the wire arrays by heating the wire cores. This results in a decreased inflow pressure, which in turn, reduces the compression ratio of the current sheet after radiative collapse. Re-absorption of emission from the reconnection layer also heats the plasma upstream of the layer, resulting in a higher temperature compared to that in the local loss simulation. The effects of optical depth on magnetic reconnection may be important in astrophysical scenarios, and this will be the subject of future study.

In order to more closely predict the dynamics of the actual experiment, we simulate a $\qty{36}{\milli \meter}$-tall load with $P_{1/3}$ multi-group radiation transport in 3D geometry. The dynamics of the 3D simulation, as described in \S \ref{sec:three-dimensionaleffects}, qualitatively reproduce those of the 2D case, exhibiting a radiative collapse process that results in decreased layer temperature and increased compression of the layer. The 3D simulation also shows strong kinking of flux ropes, with helical perturbations resembling the $m=1$ MHD kink mode, as discussed in \S \ref{sec:three_dimen_discussion}. A comparison of the MHD time with the cooling time indicates that we expect MHD instabilities to dominate right after layer formation, while cooling effects become more important later in time when radiative losses exceed the rates of Ohmic and compressional heating in the layer. The interplay of cooling and MHD instabilities provides an exciting avenue for future investigation.

The findings in this paper provide computational and theoretical evidence for rich phenomena occurring in reconnection layers with strong radiative cooling, and in particular, the role of plasmoids in localizing the radiative emission, the behavior of these plasmoids in a layer undergoing radiative collapse, and the coupling between tearing, kink, and cooling instabilities in three dimensions. 
This paper lays the groundwork for the design and interpretation of pulsed-power-driven reconnection experiments in a radiatively-cooled regime, which remains almost entirely unexplored in the laboratory. 
We therefore expect the MARZ (Magnetic Reconnection on Z) experiments to provide key insights into magnetic reconnection in this radiatively cooled regime, and the generation of high-energy emission in astrophysical systems.  
This experimental data will further augment the computational capabilities of radiation (magneto-) hydrodynamics and atomic modeling codes routinely employed in HED plasmas.

\section{Acknowledgements}

The authors would like to thank Nuno Loureiro for valuable discussions on radiative cooling of current sheets. This work was funded in part by NSF and NNSA under grant no. PHY2108050, and supported by the U.S. Department of Energy (DOE) under Award Nos. DE-SC0020434, DE-NA0003764, DE-F03-02NA00057, DE-SC-0001063, and DE-NA0003868, and the Engineering and Physical Sciences Research Council (EPSRC) under Grant No. EP/N013379/1. The simulations presented in this paper were performed on the MIT-PSFC partition of the Engaging cluster at the MGHPCC facility (www.mghpcc.org) which was funded by DOE grant no. DE-FG02-91-ER54109. This work also used the ARCHER2 UK National Supercomputing Service (https://www.archer2.ac.uk). RD acknowledges support from the MIT MathWorks and the MIT College of Engineering Exponent fellowships. DAU gratefully acknowledges support from NASA grants 80NSSC20K0545 and 80NSSC22K0828. Sandia National Laboratories is a multimission laboratory managed and operated by National Technology and Engineering Solutions of Sandia, LLC, a wholly owned subsidiary of Honeywell International Inc., for the U.S. Department of Energy’s National Nuclear Security Administration under contract DE-NA0003525. This paper describes objective technical results and analysis. Any subjective views or opinions that might be expressed in the paper do not necessarily represent the views of the U.S. Department of Energy or the United States Government.

\section{Declaration of Interests}

The authors have no conflicts of interest to disclose.

\section{Data Availability}

The data that support the findings of this study are available from the corresponding author upon reasonable request.

\bibliographystyle{jpp}
\bibliography{marz_simulations.bib}

\end{document}